\documentclass[12pt]{article}\usepackage[hyperfootnotes=false]{hyperref}
\pdfoutput=1
   \usepackage{epsfig}
      \usepackage{amsmath}
      \usepackage{amssymb}
  \usepackage{graphicx}
  \newlength{\abstractwidth}
  \renewcommand{\thefootnote}{\fnsymbol{footnote}}
  \renewcommand{\thanks}[1]{\footnote{#1}} 
  \newcommand{\starttext}{
  \setcounter{footnote}{0}
  \renewcommand{\thefootnote}{\arabic{footnote}}}
  \renewcommand{\theequation}{\thesection.\arabic{equation}}
  \newcommand{\be}{\begin{equation}}
  \newcommand{\bea}{\begin{eqnarray}}
  \newcommand{\eea}{\end{eqnarray}}
  \newcommand{\beq}{\begin{equation}}
  \newcommand{\ee}{\end{equation}}
  \newcommand{\eeq}{\end{equation}}

  \def\ba{\begin{eqnarray}}
  \def\ea{\end{eqnarray}}

  \def\12{{1 \over 2}}
  \def\eq{&=&}

  \def\d{\partial}

  \def\cc{cosmological constant }
  
  \def\simleq{\; \raise0.3ex\hbox{$<$\kern-0.75em
      \raise-1.1ex\hbox{$\sim$}}\; }
   \def\simgeq{\; \raise0.3ex\hbox{$>$\kern-0.75em
      \raise-1.1ex\hbox{$\sim$}}\; }

\def\cdl{Coleman De Luccia}

\def\ba{\bf{a}}

  \def\h3{{\cal{H}}_3}

\def\o3{\Omega_3}

\def\O2{\Omega_2}
\def\o{\omega}
 \def\d2{$dS_{2+1}$}

 \def\bi{\begin{itemize}}
  \def\ei{\end{itemize}}

\begin{document}
  \renewcommand{\theequation}{\thesection.\arabic{equation}}
  \begin{titlepage}
  \rightline{SU-ITP-11/47}
  \bigskip

  \bigskip\bigskip\bigskip\bigskip

    \centerline{\Large \bf {Eternal Symmetree}}
    \bigskip

  \bigskip \bigskip

  \bigskip\bigskip
  \bigskip\bigskip

  \begin{center}
  {{Daniel Harlow, Stephen H. Shenker, Douglas Stanford, Leonard Susskind}}
  \bigskip

\bigskip
Stanford Institute for Theoretical Physics and  Department of Physics, Stanford University\\
Stanford, CA 94305-4060, USA \\

\vspace{2cm}
  \end{center}


  \begin{abstract}


In this paper we introduce a simple discrete stochastic model of eternal inflation that shares many of the most important features of the continuum theory as it is now understood. The model allows us to construct a multiverse and rigorously analyze its properties.
Although  simple and easy to solve, it has a rich mathematical structure underlying it. Despite the discreteness of the space-time the theory exhibits an unexpected non-perturbative analog of conformal symmetry that acts on the  boundary of the geometry.
The symmetry is rooted in the mathematical properties of trees, $p$-adic numbers, and ultrametric spaces; and in the physical property of detailed balance.
We provide self-contained elementary explanations of the unfamiliar mathematical concepts,  which have have also appeared in the study of the $p$-adic string.

The symmetry acts on a huge collection of  very low dimensional ``multiverse fields" that are not associated with the usual perturbative degrees of freedom. They are connected with the late-time statistical distribution of bubble-universes in the multiverse.

The conformal symmetry which acts on the multiverse fields
is broken by the existence of terminal decays---to hats or crunches---but in a particularly simple way.  We interpret this symmetry breaking as giving rise to an arrow of time.

The model is used to calculate statistical correlations at late time and to discuss the measure problem.
We show that the natural cutoff in the model is the  analog of the so-called light-cone-time cutoff.
Applying the model to the problem of the cosmological constant, we find agreement with earlier work.

 \medskip
  \noindent
  \end{abstract}

  \end{titlepage}
  \starttext \baselineskip=17.63pt \setcounter{footnote}{0}


\tableofcontents

\setcounter{equation}{0}
\section{Introduction}

Our purpose in this paper is to present a  model of eternal inflation
which is simple enough to be completely tractable, but which has enough of the structure of the real thing that we can address some of the hard problems. The model is a stripped down version of a generalization of the Mandelbrot percolation model that was first applied to eternal inflation in \cite{Guth:1982pn},  has been used in a number of subsequent studies, notably \cite{Winitzki:2005ya}, and was employed in a recent global discussion of the phase structure  \cite{Sekino:2010vc}.

 Before we describe the model we will make some remarks of a general nature that the reader may want to come back to after reading the rest of the paper.
Let's consider a statistical system that is so simple that at first sight it seems to have nothing that could possibly interest us. The system consists of an infinite number of disconnected points. The points start out being colored red. We then go through the collection, and with probability $\gamma$ we re-color each point black. The decision as to whether we leave the points red or color them black is done independently for each point. What can we say about the system?

First of all we know that if, in an unbiased way, we take a large but finite sample of $N$ points, a fraction $\gamma$ will be black. We also know that there will be a statistical dispersion of order $\sqrt{N}.$ In fact we know the entire probability distribution for every subset of points. All of these statistical quantities will be smooth functions of $\gamma:$ there will be no phase transitions even in the infinite limit.
But now let's take the same model except that we place the points on a two-dimensional square lattice. The fact that we did so will make no difference to the questions we previously discussed. It is still true that if we pick any subset $N,$ the average number of black points will be $\gamma N.$ But now there are some new questions we can ask. For example, what is the distribution of sizes of percolation clusters? At what value of $\gamma$ does the system undergo a percolation transition? How many different percolation phases does the system have and what is the order of the transitions between them? To go a step further, take the points and put them on a $D$-dimensional cubic lattice. The same kinds of percolation questions can be asked but the answers are $D$-dependent even though it is exactly the same set of points with exactly the same statistical rule for coloring them. By adding the structural relations implicit in the lattice we have made a very dull model interesting. Notice that the original statistical model had nothing in it which told us whether points are distributed on a lattice; or, if they are,  what is
the  dimension of the lattice.

The model of eternal inflation we are going to study is in some ways similar to this simple example---not the interesting lattice version, but rather the dull  version---which has so little structure that one cannot say how many dimensions it represents. The features that it exhibits are those that are common to every dimensionality. We will argue though, that as simplistic as the stripped down model may be, it still has enough complexity to address many of the subtle problems of eternal inflation.  Conformal symmetry is still present in the stripped down version, and non-terminal bubbles don't disrupt it. Terminal bubbles break the symmetry, but in a very simple way. And most notably, crucial elements for studying the measure problem have not been eliminated by the stripping down.

The reason that we can profitably consider so unstructured a model is that the large scale properties---super-horizon properties---of eternal inflation are dominated by the fact that different causal patches have passed out of causal contact and do not interact with each other. That is the main feature of the model we study; it is a model of causal patches that reproduce and fall out of contact.

The model described in this paper shares many features with the model proposed
by Freivogel and Kleban  in
\cite{Freivogel:2009rf}.   Other notable work on this subject includes \cite{Garriga:2009hy}.

In the remainder of the paper we give a systematic discussion of the model.   After concluding we present two appendices.  The first reviews the properties of eternal inflation that parallel the cellular model, and the second gives a few brief remarks on some aspects of dS/CFT and FRW/CFT motivated by this model.

\setcounter{equation}{0}
\section{The Cellular Model}
\subsection{The Model}

We will start with a stochastic \cite{Starobinsky:1986fx}  model which seems  complicated enough to have a definite dimensionality.
We began studying this model as an approximate discretization of the standard bubble-nucleation \cite{Coleman:1980aw} picture of eternal inflation.   However, we soon came to realize that the cellular model is not so much an approximation to the continuum theory,  as it is a mathematically distinct structure which exhibits remarkable parallels with eternal inflation.

We first describe a version of the model assuming a landscape containing only
de Sitter vacua. We do not derive the landscape from string theory or from a model of scalar fields: we simply postulate a discrete collection of vacua labeled by an index $n$ called the \it color \rm of the vacuum. In addition to the color label, each vacuum has a \cc \ proportional to the square of its expansion rate $H_n.$ We will also attribute to it an entropy,
\be
S_n \sim {1 \over  H_n^2 G}.
\label{Sn}
\ee
The entropy counts the number of microstates that are lumped into a single macroscopic state $n.$
Later we will  include
 terminal vacua
with zero or negative \cc.

The model can be constructed in any number of spatial dimensions but for definiteness we will work with 3+1 dimensional spacetime. There are two versions of the model which we will call the compact and noncompact cases. The compact case begins with a single cell, while the non-compact case begins with an infinite cubic lattice of cells.
The two cases represent the global slicing of de Sitter space by compact spatial 3-spheres, and the non-compact flat-space slicing of de Sitter space.
In either case we   pick an initial condition \cite{Garriga:2006hw} in which each cell is painted with a color. In the non-compact case the colors can vary from place to place but for simplicity we take the initial state to be uniformly colored.

Inflation  is represented by dividing the cells in half, along each direction of space, so that every cell is replaced by  $8 = 2^{3}$ smaller cubes without changing the color. This represents a single 2-folding (we use the term 2-folding in the same sense as e-folding) of inflation. If we endlessly  repeat the doubling procedure  without changing the colors of the cells, the process models the exponential expansion of de Sitter space. However, the actual procedure defining the model is more complicated.

\begin{figure}[h]
\begin{center}
\includegraphics[scale=.6]{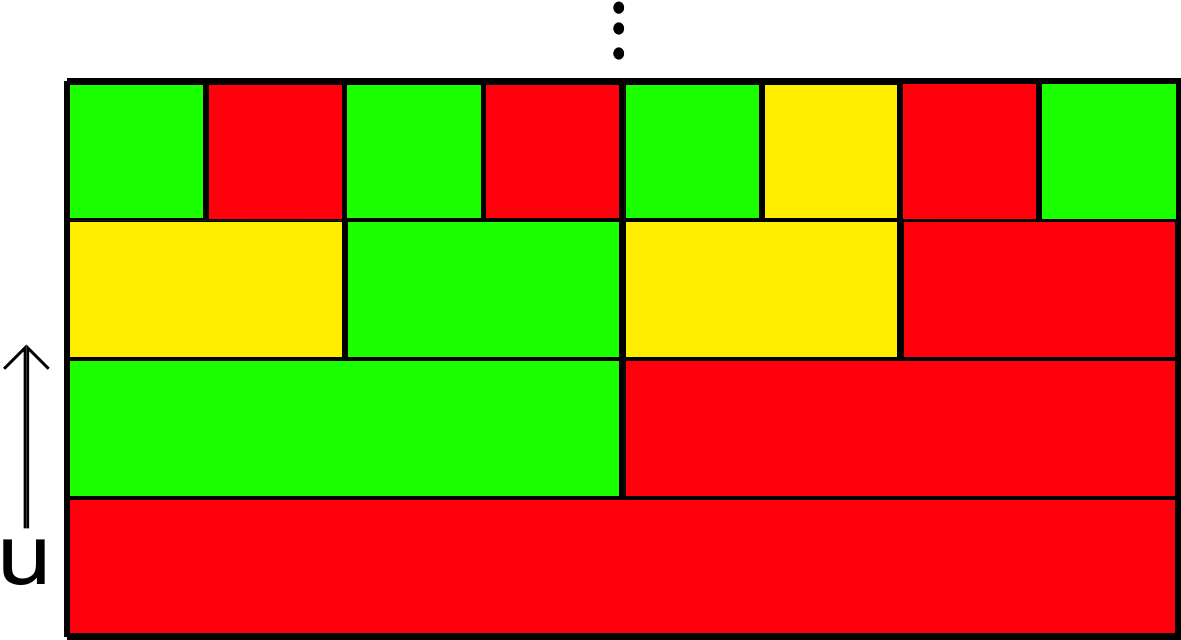}
\caption{Three steps of the 1+1 dimensional cellular model.  Note that transitions in both directions are allowed.}
\label{fig:bubbles}
\end{center}
\end{figure}

After doubling the lattice (multiplying the number of cells by $2^3$), color the boxes by going through the finer lattice and, with probability $\gamma_{nm} \ll 1$, repainting each $m$-colored cell with color-$n$. The symbol $\gamma_{nm}$ represents the rate of nucleation of  bubbles of type $n$ in a vacuum of type $m$.
This procedure is iterated ad infinitum in the obvious way. At each step the cells are divided into $2^3$ equal cells, which are then re-colored with probabilities $\gamma_{nm}$.  This procedure is illustrated for the 1+1 dimensional case in figure \eqref{fig:bubbles}.

After $u$ steps the number of cells is $N(u) =2^{3u}$ and the number of cells of color $n$ is $N_n(u).$ The fraction of cells with color $n$ can be thought of as the probability that a $u^{th}$ generation cell has color $n,$
 \be
 P_n(u) \equiv {N_n(u) \over N(u)} = {N_n(u) \over 2^{3u}}.
 \ee

The probabilities  $ P_n(u)$ are governed by rate equations similar to those introduced for bubble nucleation in eternal inflation by Garriga, Schwartz-Perlov, Vilenkin and Winitzki \cite{Garriga:1997ef}.  In going from step $u$ to $u+1$ the probabilities change according to
 \be
 P_m(u+1) = P_m(u) -\sum_n \gamma_{nm} P_m(u) + \sum_n  \gamma_{mn} P_n(u).
 \label{rate equation in comp form}
 \ee
The negative term on the right side represents the depletion of the $m^{th}$ color by transitions to other colors. The final term is the increase due to transitions from other colors. Note that \eqref{rate equation in comp form} guarantees probability conservation:
$$\sum_m  (P_m(u+1)-P_m(u)) = 0.  $$
We can also write the equations in matrix form. Let $P$ be a column vector with entries $P_n.$
\be
 P(u+1) = G P(u)
\label{rate eq in vector form}
\ee
with the matrix $G$ having the form
\be
G_{mn} = \delta_{mn}-\sum_r \gamma_{rm}\delta_{mn} + \gamma_{mn}.
\label{G in terms of gamma}
\ee

 \subsection{Detailed Balance}

 Although the rate equations make sense without any restriction on the symmetry of the matrix $\gamma_{nm},$ in what follows the property of \it detailed balance \rm will play an important role. One can justify detailed balance in a number of ways, the most relevant one for our considerations being that it is true for \cdl \ tunneling rates.
Detailed balance says that $\gamma_{nm}$ and $\gamma_{mn}$  differ only in that the transition rate is larger in the direction that increases the number of microstates.  More precisely, it assumes that the transition rates between between microstates are symmetric.  The detailed balance condition is
  \be
 {\gamma_{nm} \over \gamma_{mn}}= e^{S_n -S_m}.
 \ee
This rule can be expressed in terms of a real symmetric matrix $M_{mn} = M_{nm}$:
 \be
\gamma_{nm} = M_{nm}e^{S_n}.
\label{detailed balance}
\ee

Note that according to this rule transitions between vacua can proceed in both directions---from smaller to larger $S,$ or from larger to smaller---
although the probability for a transition that decreases the entropy is much smaller than then the transition that goes in the other direction.  The motivation for allowing upward transitions comes from considering the thermodynamics of a static patch of de Sitter space \cite{Dyson:2002pf,Brown:2007sd,Aguirre:2011ac}.  

The matrix $G$ in \eqref{G in terms of gamma} is not symmetric but it does have a complete set of eigenvectors with real eigenvalues. To see this
define\footnote{This was pointed out to us by Yasuhiro Sekino.} a diagonal matrix $Z$
\be
Z_{mn}=\delta_{mn}e^{S_n/2}
\ee
and a new matrix $S$ by
\be
S=Z^{-1}GZ.
\ee
Plugging in the value of \eqref{G in terms of gamma} and \eqref{detailed balance} one sees that the matrix $S$ is symmetric. It thus has a complete orthonormal set of eigenvectors, which we denote by
\be
\sum_n S_{mn}(I)_n=\lambda_I(I)_m.
\ee
These satisfy
\be
(I)\cdot (J)=\sum_m (I)_m  \ (J)_m=\delta_{IJ}
\ee
and
\be
\sum_I (I)_m  \ (I)_n = \delta_{mn}.
\label{orth n}
\ee
The eigenvectors of $G$ are then given by
\be
\label{Phi}
P^{\{I\}}_m=e^{S_m/2}(I)_m.
\ee
\subsection{Statistical Equilibrium}

The following facts are easy to prove.
\begin{itemize}
\item The sum of the $P_m$ is conserved.  To interpret $P_m$ as  a probability we normalize the sum to 1.
\item There is exactly one eigenvector with unit eigenvalue. We call it $(0)_m$
\item All other eigenvalues have magnitude less than one.
\item The eigenvector $(0)_m$ has the form
\be
(0)_m=\frac{e^{S_m/2}}{\sqrt{\sum_n e^{S_n}}}.
\label{null eigenvector Phi}
\ee
\end{itemize}

The interpretation of \eqref{null eigenvector Phi} is most obvious when it
is expressed in terms of the
probabilities $P_m$:
\be
P^{\{0\}}_m = \frac{e^{S_m}}{\sum_n e^{S_n}}.
\label{null eigenvector P}
\ee

Since all other eigenvalues have magnitude less than one, a generic initial condition
for $P_n$ will evolve to become asymptotically proportional to \eqref{null eigenvector P}. In other words the system evolves to an equilibrium fixed-point, in which the population of vacua is simply proportional to the number of microstates of each vacuum.
As we will see this result is entirely expected from a local perspective.

Let us consider the approach to the fixed-point. The  transient behavior is determined by  the remaining eigenvalues of $S$ which are all less than one in magnitude. In particular the  largest eigenvalue not equal to one is called $\lambda_1.$ It determines the leading transient. The corresponding eigenvector is $P^{\{1\}}_m=e^{S_m/2}(1)_m$. The leading transient will have the form
\be
P=P^0_m + c \lambda_1^u P^{\{1\}}_m.
\label{transient behavior}
\ee

Initial conditions are easy to formulate in terms of the eigenvectors $P^{\{I\}}_m$. The general solution of the rate equations is
\be
P_m = \sum_I c_I P^{\{I\}}_m \lambda_I^u.
\label{transient in continuous time}
\ee

\subsection{Terminals and the Dominant Vacuum}
Terminal vacua play a very important role in eternal inflation. A terminal vacuum is one from which no further transitions take place. Vacua with vanishing \cc \ are terminal. Vacua with negative \cc \ are usually assumed to also be terminal.

When there are terminal vacua the situation is somewhat different. For simplicity suppose there is one terminal vacuum labeled by $m=0$.\footnote{It is trivial to include more than one terminal, one just changes $\gamma_{m}$ below to $\sum_i \gamma_{im}$ where $i$ runs over all terminals.} By definition the rate for a terminal to make a transition to any other vacuum is zero. Therefore  $\gamma_{n0} =0$ for all $n$. Formally this is what would happen if $S_0$ were infinite. In that case the eigenvector $P^{\{0\}}$ would be trivial; it would have support only on the terminal entry. The equilibrium fixed-point would degenerate so that only the terminal vacuum would have non-vanishing weight.

We can eliminate the terminal vacuum from the rate equation altogether.  Defining $\gamma_m \equiv \gamma_{0m}$, the population of non-terminal vacua is controlled by the rate equation
\be
 P_m(u+1) =P_m(u) -\left(\sum_n \gamma_{nm} +\gamma_{m}  \right)P_m(u) + \sum_n  \gamma_{mn} P_n(u).
 \label{nonterminal rate equation}
 \ee
Here $m,n$ are non-terminal and $\gamma_m$ is the transition rate from color ${m}$ to the terminal vacuum. These equations no longer satisfy probability conservation for the simple reason that probability leaks into the terminal vacuum.  The matrix $G$ is now given by
\be
G_{mn}=\gamma_{mn}+\delta_{mn}(1-\gamma_n-\sum_r\gamma_{rn}).
\ee
The matrix $S=Z^{-1}GZ$ is still symmetric and its eigenvectors still form an orthonormal basis, now with all eigenvalues less than one in magnitude.

The largest eigenvalue of this $G$ determines the asymptotic late-time population of non-terminal vacua.  Call it and its corresponding eigenvector $\lambda_D$ and $(D)_m.$ We also define $P^{\{D\}}_m$ through equation \eqref{Phi}.  The statistical ensemble of vacua with relative probabilities  $P^{\{D\}}_m$ is called { \it  the dominant eigenvector}  \cite{Garriga:1997ef}.

It is widely expected that decay rates $\gamma$ are very dissimilar to one another since they are exponentials of \cdl \ instanton actions. Generically this  leads to a situation in which one vacuum---the dominant vacuum---has a much longer lifetime that any other.  The dominant eigenvector will then have entries which are very small except for the dominant vacuum. In what follows we will \it not \rm make use of the assumption that only one vacuum dominates $P^{\{D\}}.$

\subsection{Series and Parallel}

There are two ways to follow the evolution of the model. The first is global. In the global view, at any given time the system consists of $2^{3u}$ cells which all split simultaneously with each tick of the clock, i.e., each integer step of $u.$ . The number of cells of color $n$ is $N_n(u)$. Some patches reach terminal status and stop reproducing, but eternal inflation continues.

The second way of thinking about the model is completely local and does not require any global considerations. Imagine yourself in one of the cells. When the cell splits, pick one of its descendant cells---it does not matter which---and imagine transiting into that cell. At each stage you are following a branching tree along a route from its trunk to one of its branches. If at any given time you are in a cell of color $n$ then you may stay in the same color, or with probability $\gamma_{mn}$ transition to color $m.$ Following the process locally, the probability of finding yourself in a given color satisfies the rate equation \eqref{rate equation in comp form}. In this form the evolution refers to a single causal patch, and it terminates when a terminal vacuum is reached.

The relation explained above, between the parallel and series views, is at the heart of the ``global-local duality" described in \cite{Bousso:2009mw}.

Consider, from the local perspective, the case with no terminal bubbles. This is the analog of studying de Sitter space in the causal patch formulation. What one expects is that the patch eventually comes to  thermal equilibrium. Thermal equilibrium does not mean the system is static: fluctuations continually take place between configurations, with the statistical distribution of vacua (local minima of the landscape-potential) being proportional to \eqref{null eigenvector P}.

From the local perspective the probability $P_m(u)$ is nothing but the probability for the observer to observe color $m$ at time $u$. If the observer's measurements are conditioned on $u$ being large then $P_m(u)$ will either be governed by the fixed point \eqref{null eigenvector P} if there are no terminals, or by the dominant eigenvector if there are terminals.

Of course the observer has no way of knowing how large $u$ is. If it is not large the observer will experience transients which depend on the initial condition. Such a behavior would be surprising from the global view since the overwhelming majority of branches are very high up on the tree.

\setcounter{equation}{0}
\section{The Tree-like Geometry of the Model}
\subsection{Why Trees?}

In this section we are going to strip away some superstructure of the cellular model. The superstructure is roughly the analog of adding a lattice to the simple statistical model in the introduction. Eliminating it would be the wrong thing to do if we were interested in percolation, but if our interest is only in the population-statistics of black and red points, the lattice may be  confusing. In the model of colored points there is no geometric or topological structure left after we throw away the lattice, but the stripped-down cellular model has an interesting geometry that survives. Moreover, for the questions that we will be concerned with; namely, questions of statistics and correlations of populations, the lattice superstructure is irrelevant. That's not to say that there are no questions for which it is relevant.

One of the chief advantages of eliminating the superstructure is that it obscures a
powerful symmetry of the model. The surprising exact symmetry
 closely parallels the full symmetry of de Sitter space, including the conformal symmetry of its boundary.

The important fact in stripping away the excess structure is that
two cells, which are created at the same instant, never talk to each other afterwards; they branch and evolve completely independently. This is so, not only for cells which are distant  on the lattice, but also for adjacent cells. The really important notion of proximity is not how close points are on the lattice, but how far back into the past you have to go before the two cells are found within a common ancestor cell. We discuss this in greater detail in section 5.

 Given that once cells form, they never interact, it is obvious that the structure of the model is tree-like. Tree graphs are composed of nodes and links (often called edges) and have no closed loops. The tree representing the compact version of the cellular model begins with a single link which then branches out to eight links. Each new link then branches to eight more ad infinitum.

 Once we strip away the lattice superstructure there is nothing special about the number eight:
 we may consider tree models which at each stage branch to two, three, or any number of branches. Very little of what we say depends on that number, but for one or two technical points involving the symmetry of trees, the mathematics is simplest if
  at each step an incoming branch splits into a prime number of outgoing branches.\footnote{Our computations of correlation functions can be easily generalized to arbitrary branching number, but the description of the symmetry of the tree becomes more complicated if it is not prime.}

\begin{figure}[h]
\begin{center}
\includegraphics[scale=.6]{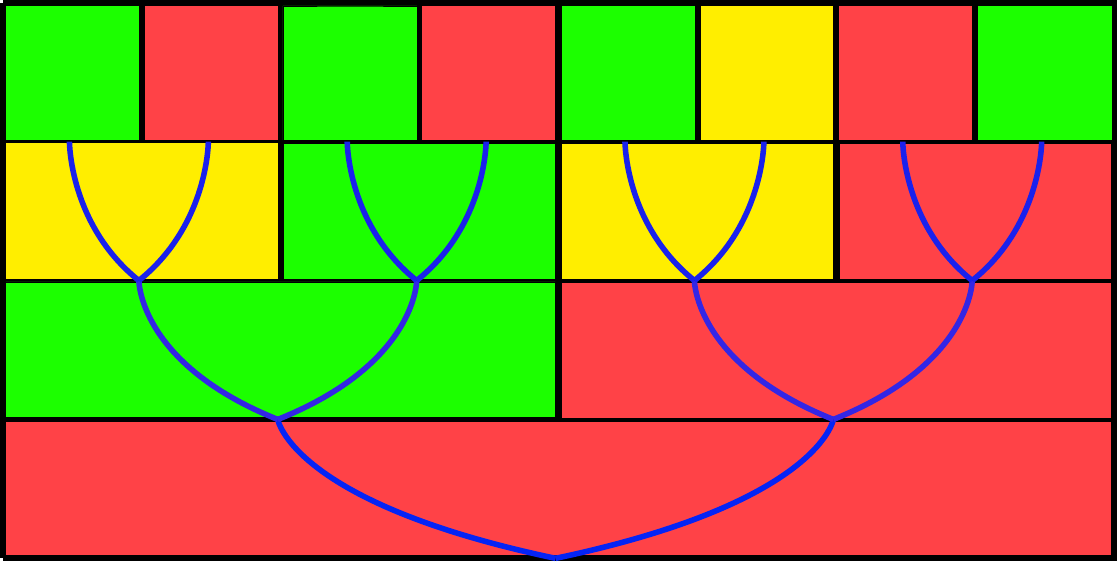}
\caption{The causal tree for three steps of the $p=2$ cellular model.}
\label{fig:bubbletree}
\end{center}
\end{figure}

  The simplest of all is a tree in which each branch splits into two branches. We will mostly consider the general prime case, often illustrating it with the binary case as in   figure \eqref{fig:bubbletree}. We will use the letter $p$ to denote the number of outgoing branches: the binary case is $p=2.$

Before discussing the tree further, let us recall some concepts about the causal structure and geometry of Lorentzian space-times and de Sitter space in particular.
The discussion is not intended to be complete.

\bi
\item

In a Lorentzian space-time there is a causal relation between any pair of points $a$ and $b.$  The point $a$ is said to be in the  future of $b$ if there is a future directed time-like or light-like trajectory from $b$ to $a.$ For simplicity we will just say time-like.
We may also say that $b$ is in the  past of $a.$ The set of points in the  future of $a$  defines  the causal future of $a.$ The set of points in the past  of $a$  defines  its causal past.
\item

Two points will be said to be out of causal contact if there is no third point in the causal future of both of them. In other words they are out of causal contact if they cannot send messages to the same point.

\item

If $a$ is in the  past of $b$ then the intersection of the causal future of $a$ and the causal past of $b$ is called the causal diamond of $a,b.$

\item

 A causal structure is homogeneous if any two points are related by a mapping of the geometry that preserves the causal structure.

\ei

De Sitter space has a particular causal structure that satisfies the following postulates:

\bi

\item
de Sitter space has a homogeneous causal structure. There are symmetries which map any point to any other point.

\item
A special feature of inflating spaces is that future directed world lines can fall out of causal contact. This is described by saying that for any point $a,$ there exist points $b$ and $c$  in the causal future of $a,$ that are out of causal contact with each other.

\ei

Now let us return to the tree geometry  of the cellular model.

\bi
\item  Every node influences all the links that grow out of it in the future direction. The set of nodes and links that can be influenced by a given node is called the \it causal future \rm of the node. It is itself a sub-tree and it plays the role of the interior of the future light-cone of the node.  

\begin{figure}[t!]
\centering
{\includegraphics[scale=0.33]{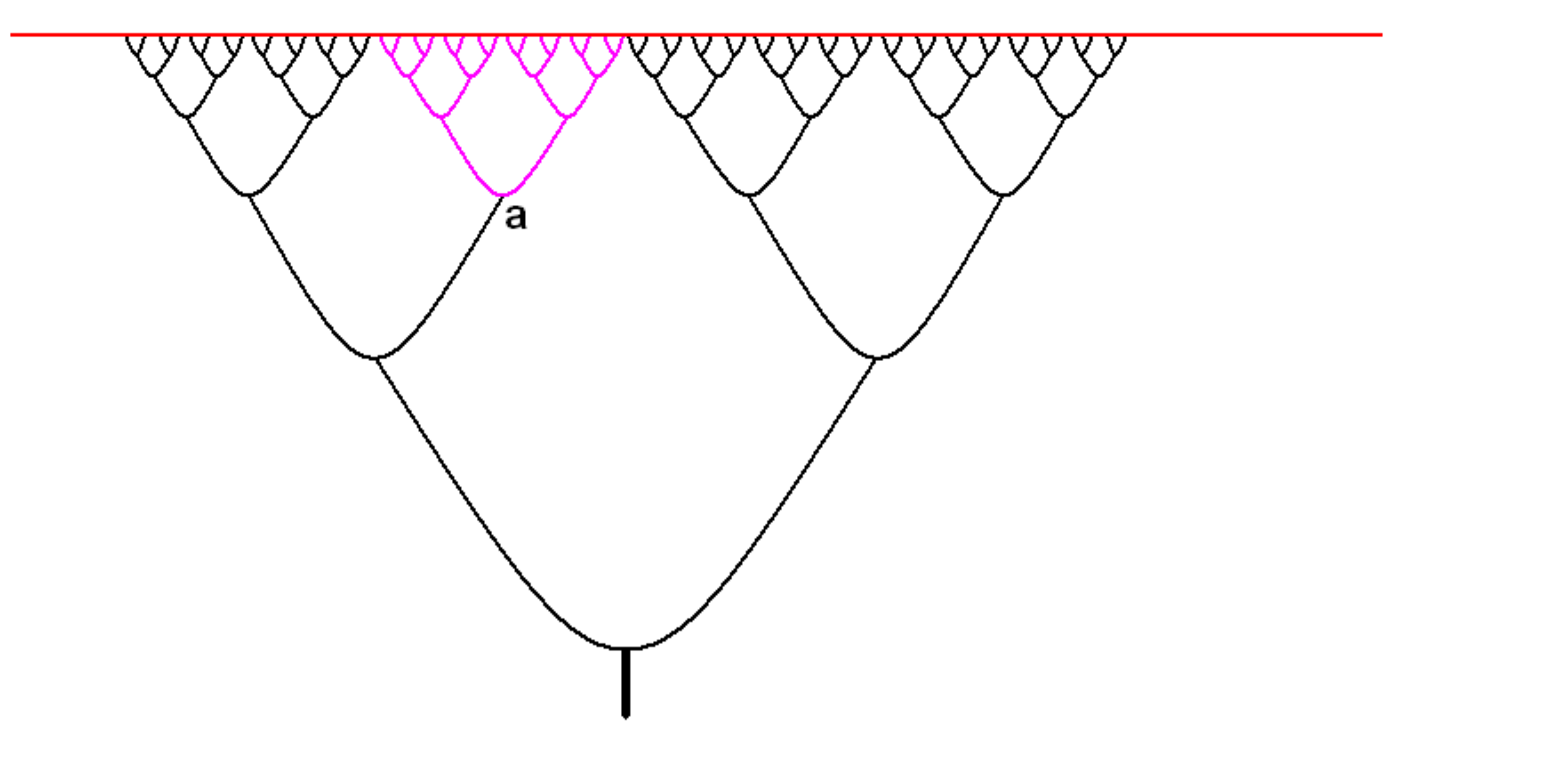}}\hspace{1mm}
{\includegraphics[scale=0.33]{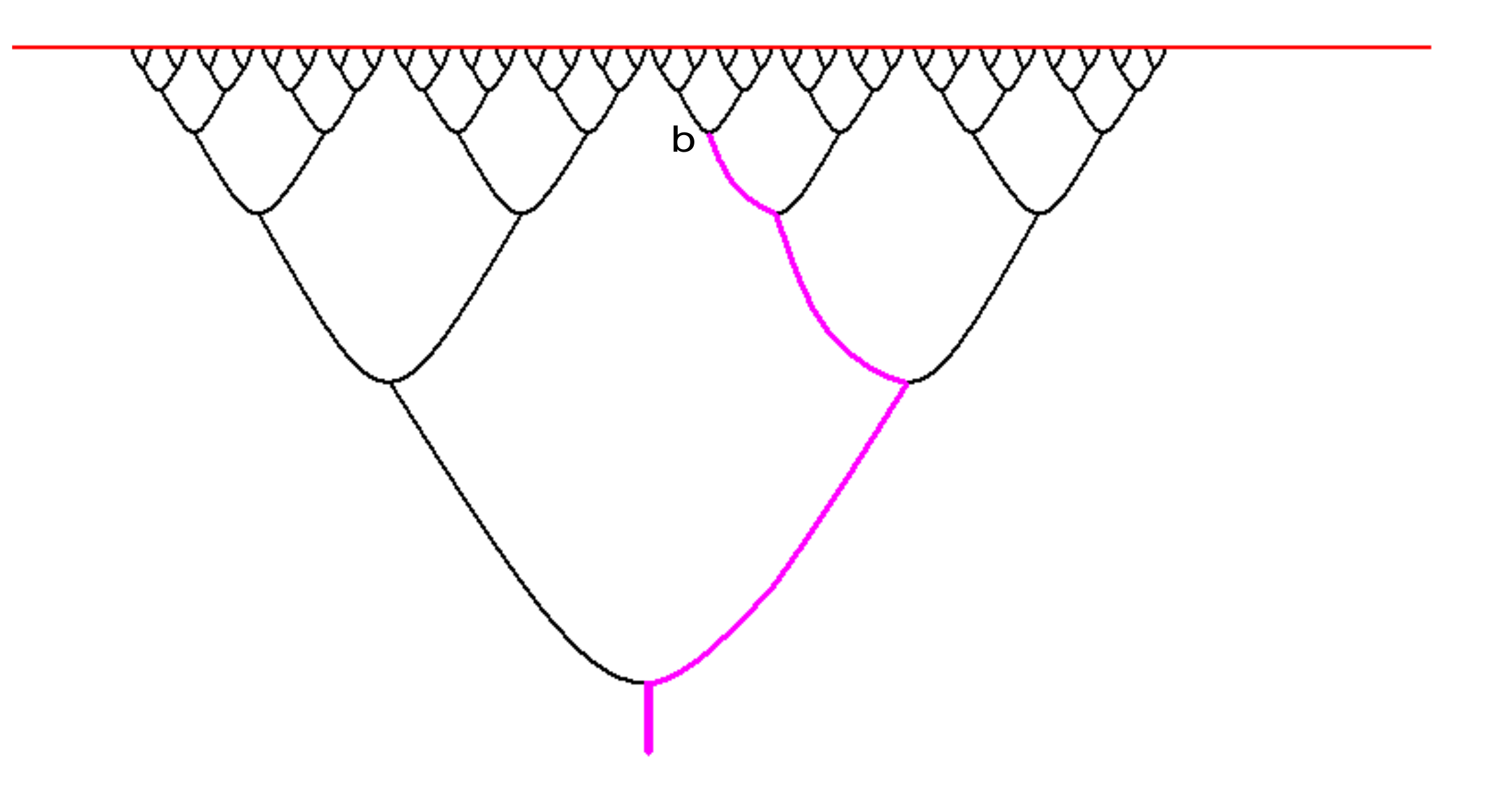}}
\caption{The causal future of $a$ and the causal past of $b$ on the $p=2$ tree.}
\label{f23}
\end{figure}

   \item The \it causal past \rm of a node $a$ is the set of all  points whose causal futures contain $a.$ In other words the  causal past  of  $a$ is the set of points that can influence $a,$ or that can be seen from $a.$  The causal past plays the role of the interior of the past light-cone of $a.$
Figure  \eqref{f23} shows the causal future and past of two points, $a$ and $b$ respectively, as pink subsets of links.

       \item Given two nodes $a$ and $b$ with $a$ in the causal future of $b,$ we define their causal diamond as the intersection of the causal future of $a$ and the causal past of $b$ 

        \item It is not obvious in what sense the tree is homogeneous and in fact it is not in the compact case.  But as we will see, the non-compact case has a large symmetry that makes it homogeneous. The symmetry will be called $p$-conformal symmetry.

            \item It is obvious by inspection that for any point $a,$ there exist points $b$ and $c$  in the causal future of $a,$ that are out of causal contact.

          \ei
          We can also introduce a metrical structure for the tree. The simplest metrical structure is to assume every edge is of the same length. We will call that notion of distance, the graph-distance.  But in the cellular model there is another, fluctuating metric. Recall that each cell is endowed with a color $n$ and Hubble constant $H_n$ which  are determined probabilistically. The Hubble constant of an edge provides a unit that allows a definition of proper time.

          \bi

       \item The proper time between two successive nodes is given by
\be
\Delta \tau = H_n^{-1}
\ee
where the $n$ is the color of the link connecting the points. More generally, if $b$ is in the causal future of $a$ then the proper time between the two nodes is
\be
\tau_{ab} = \sum_n H_n^{-1}.
\label{proper time}
\ee

          \ei

So far the tree picture has been described assuming no terminals. The existence of terminals is a straightforward modification. If a terminal branch grows out of the tree it simply gets pruned after one link.
The main effect of the terminals is to make the total number of branches grow slightly slower than $p^u.$
\begin{figure}[ht]
\begin{center}
\includegraphics[scale=.3]{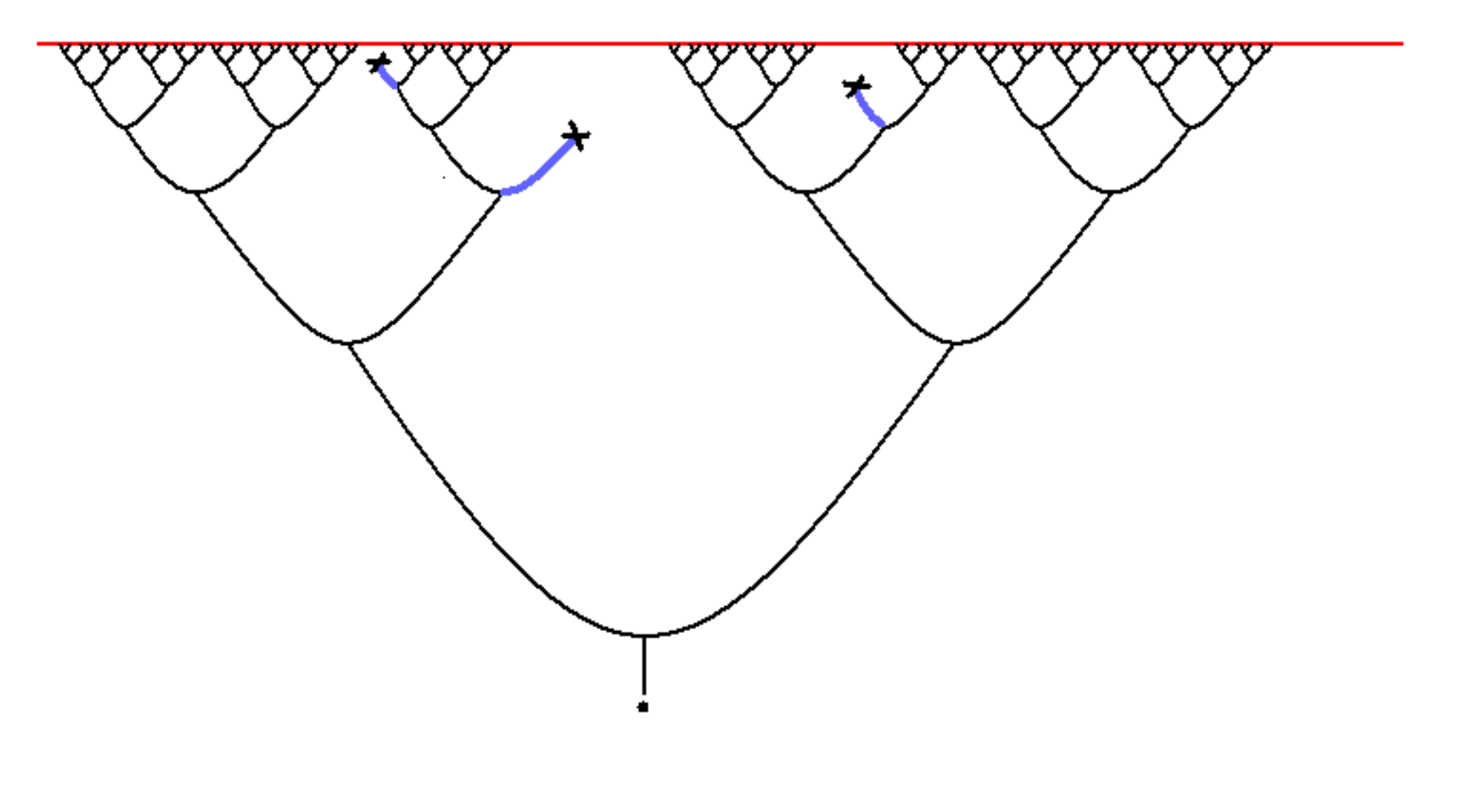}
\caption{The pruned tree of a landscape with terminals.}
\label{f6}
\end{center}
\end{figure}
\subsection{The Future Boundary}

In the theory of de Sitter space, and also eternal inflation, the future boundary plays a very important role \cite{Witten:2001kn,Strominger:2001pn,Maldacena:2002vr}. The future boundary is not a part of the original bulk geometry. It is a set of added points that close the geometry. It may be defined in the following way. Consider the set of infinitely extended time-like world-lines. These world-lines form equivalence classes. We will say that two such world-lines are in the same class if they never fall out of causal contact. In other words for any point on one world-line some part of the other world-line is in its causal future.  Each equivalence class defines a single point on the future boundary of de Sitter space. Note that by definition, any two distinct boundary points are out of causal contact.

One can define the causal past of a boundary point in an obvious way. The causal past of a boundary point is called a causal patch.  Now let us turn to the tree.

\begin{figure}[ht]
\begin{center}
\includegraphics[scale=.3]{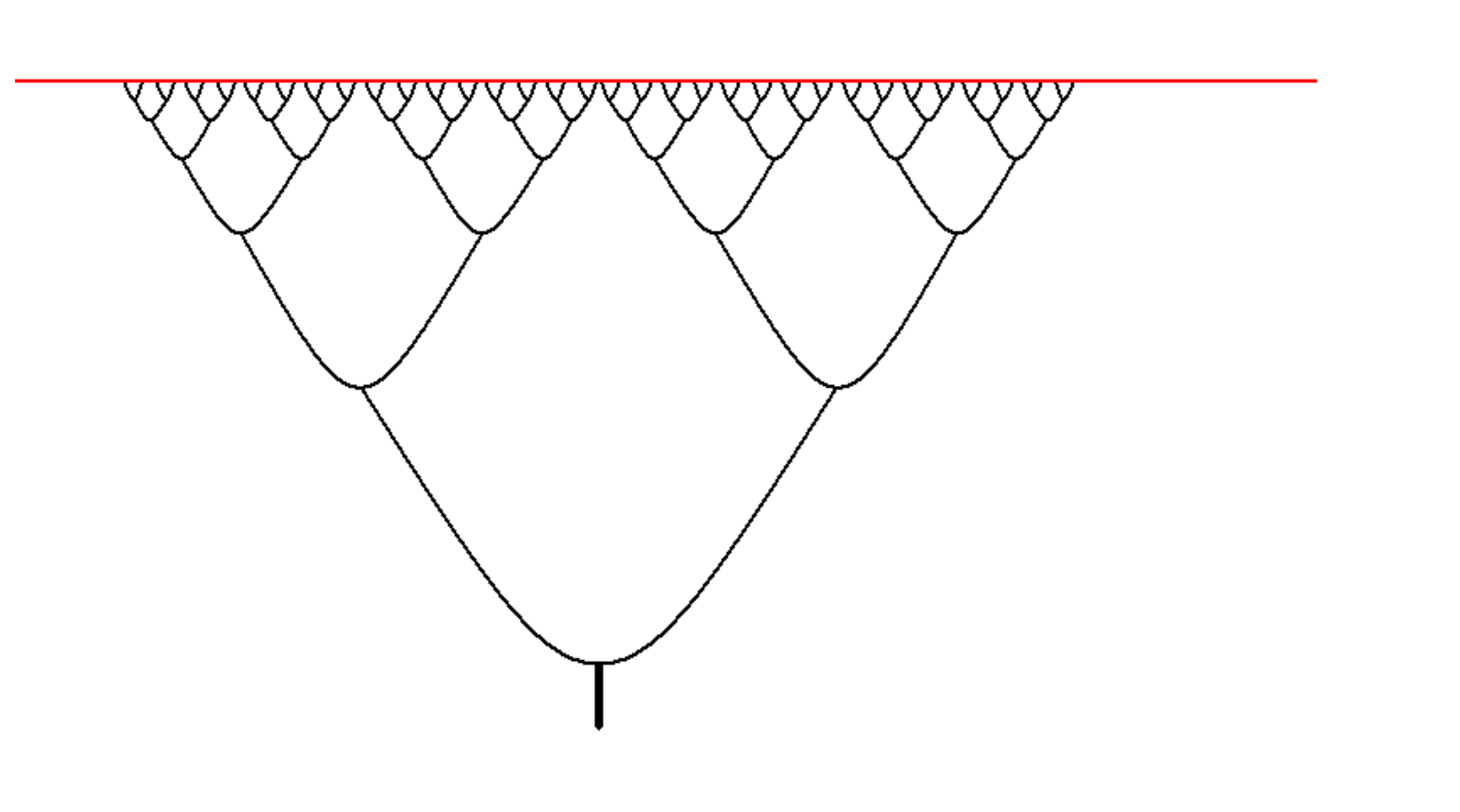}
\caption{The future boundary of the infinite tree for the $p=2$ model.}
\label{f1}
\end{center}
\end{figure}

The red line at the top of the infinite tree in Figure \eqref{f1} is the future boundary of the tree. It is defined in much the same way as in the continuum case. Take any infinite future directed sequence of links. Two distinct sequences will always fall out of causal contact, so each equivalence class contains only a single world-line in the tree case. The endpoint of each infinite world-line defines a point on the boundary. The world-line itself defines a causal patch.

One can regulate the tree by terminating it after a given number
of steps, i.e., after $u_0$ units of discrete time. The regulated version of the boundary
is composed of a set of points which grows exponentially with $u_0.$
The regulator time $u_0$ eventually tends to $\infty.$

The boundary has its own geometry which is not the geometry inherited from the bulk.
Consider two points $x, \ y,$ on the boundary. The \it boundary-distance \rm between $x$ and  $y$  is defined in what may appear to be an odd way. We consider the unique path from each point $x, \ y,$   that proceeds toward the past. Eventually those paths will intersect at some value of the discrete time $u_i(x,y)$ (the index $i$ indicates intersection). The distance between the points $x,y$ is be defined to be
   \be
   |x-y|_p =p^{-u_i(x,y)}.
    \label{u-metric}
    \ee

At the moment the left-hand side of this equation is just a symbol, but we will see below that  it is $p$-adic distance defined on the boundary.    Note that in this form there is no reference to the regulator $u_0$.

The distance defined by \eqref{u-metric} has an unusual property. Pick any three points $x,y,z$ on the regulated boundary. It is always the case that two of the three distances
$  |x-y|_p, \  |x-z|_p, \   |z-y|_p$ are equal. The third one is either shorter than, or equal to the other two. In other words every triangle is either equilateral or ``tall-isosceles". The term for geometries with this property is \it ultrametric \rm  \cite{Rammal:1986zz}. This is equivalent to a strong form of the triangle inequality:
\be
|x-y|_p \leq  \max \ \left\{ \ |x-z|_p, \ |z-y|_p \ \right\}
\ee

It should be noted that the ultrametric boundary distance has nothing to do with the bulk metric  defined by \eqref{proper time}. The bulk metric can be thought of as fluctuating since the colors are determined by probabilistic rules, while the boundary ultrametric is a fixed property of the tree.\footnote{This structure should be closely related to the ultrametric structure in perturbative de Sitter fluctuations found by Anninos and Denef in \cite{dio}.}

\subsection{The p-adic  Boundary }

The boundary of de Sitter space is an ordinary Euclidean geometry. The boundary of the tree is less familiar and is described in terms of the $p$-adic numbers.
If the noncompact tree is defined so that each incoming branch splits into $p$ outgoing branches then the boundary of the tree is the space of $p$-adic numbers which we will define shortly\footnote{If the number of outgoing branches is not equal to a prime, then much of the discussion is unchanged if we substitute the N-adic numbers.  The $p$-adic number systems are special however in that they are algebraic fields.  This is not true for arbitrary $N$, for example in the 10-adic numbers one can find two nonzero elements  whose product is zero.  These elements cannot have multiplicative inverses, so it is not always possible to divide one N-adic number by another.  That such division is possible for $p$-adics will be important in our discussion of symmetry below.}.

The $p$-adic
 numbers (and also the N-adics) are well known to have an ultrametric structure. In this section we will give an elementary account of $p$-adic numbers and their relation to trees.   For an especially clear review  we recommend \cite{Zabrodin:1989cd}. These concepts have been used in the study of the $p$-adic string \cite{Freund:1987kt}. Here the string  world sheet is the tree \cite{Zabrodin:1989cd}.  On a formal level  $p$-adic string correlators are a special case of those discussed here.

 We first consider the case in which the tree grows out of a single branch as in Figure \eqref{f1}.  This case is simpler because only the $p$-adic \textit{integers}, denoted $\mathbb{Z}_p$, are needed to describe the boundary.  Our description will often specialize to $p=2$ but the generalizations are straightforward. We will write numbers in base $p.$ Begin at the decimal point and work to the left as in the usual theory of integers. For example
$100101100.$ is the binary expansion of $300.$ Now, unlike the ordinary integers, $p$-adic integers are defined as being unending sequences of this type: unending to the left! Thus $300$ is written $$...00000100101100.$$ As long as the non-zero part of the string is finite the number is an ordinary integer. But in the theory of $p$-adic numbers  we  allow arbitrary  infinite strings, most of which would have no meaning in ordinary arithmetic.

We may formally represent the $p$-adic integers as infinite sums of the form
\be
x= \sum_{u = 0}^{\infty} x(u)p^u
\ee
where $u$ runs over nonnegative integers and the coefficients $x(u)$ are integers chosen from $\{0, \ 1,... \ p-1\}.$ In other words the $p$-adic integers are just the usual expansion of integers in base p, but with one difference: they are allowed to go on forever. Of course most of these numbers would be infinite with the usual notions of magnitude but we won't let that stop us.  To keep the discussion as concrete as possible we will occasionally introduce a cutoff by bounding $u.$ The maximum value of $u$ is called $u_0.$ In the end, $u_0$ will be allowed to tend to infinity.

\begin{figure} \begin{center} \includegraphics[scale=.4]{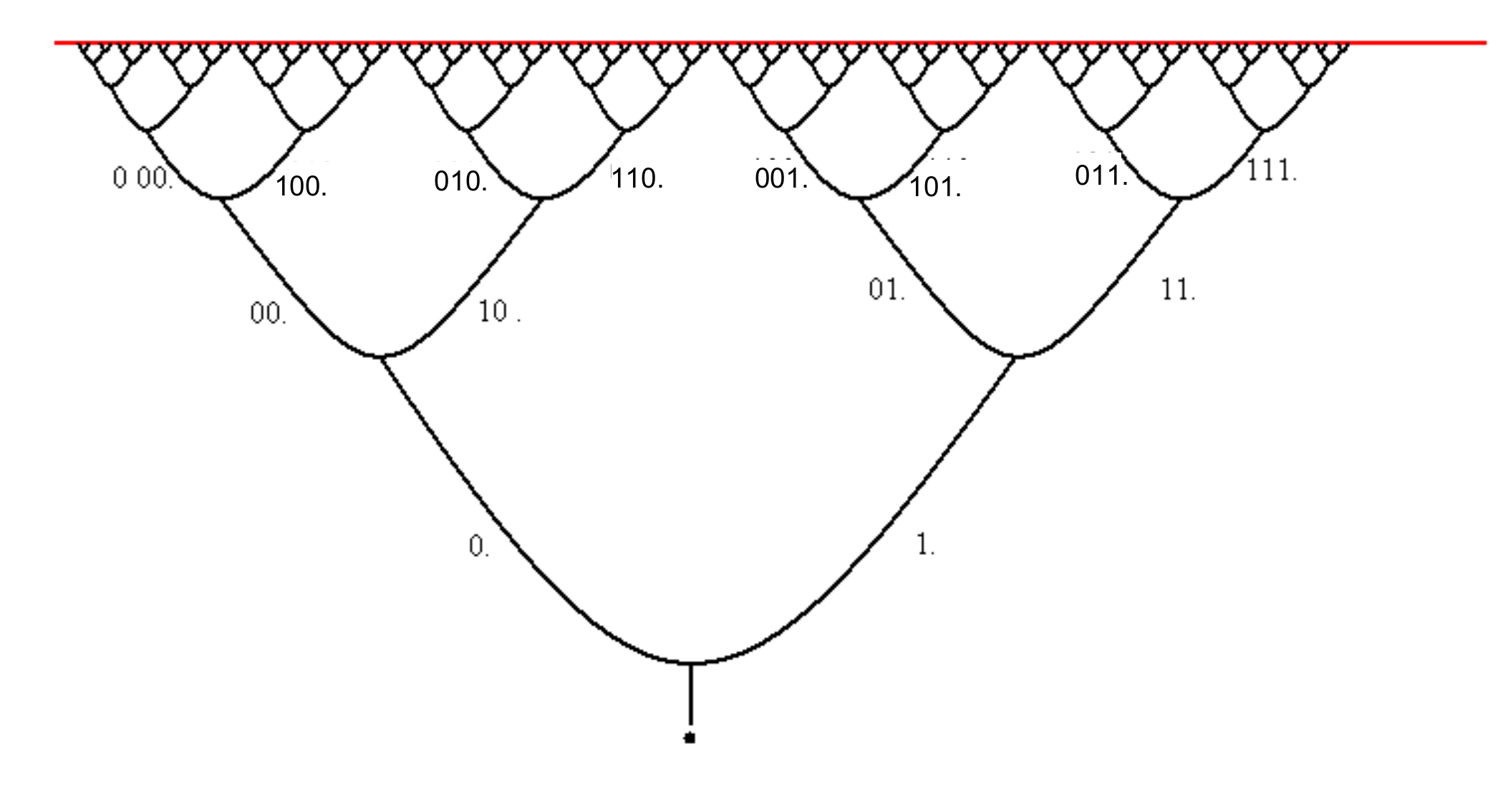} \caption{The tree structure of the 2-adic integers.  Note that although elements appear multiple times in the tree, for example we have written zero three times, they only appear once at a given level and the boundary counts each $p$-adic integer only once.} \label{f5} \end{center} \end{figure}

The $p$-adic integers can be organized in terms of a tree. The root of the tree is the decimal point at $u=0.$ Next we move up to $u=1$ where we create the first branching by writing either $0,1,...,p-1$ to the left of the decimal point.  For the illustrative case of the 2-adic numbers the two branches are $0.$ and $1.$ We iterate the procedure: at $u=2$ each branch splits by writing either a $0$ or $1$ to the left. Thus the branch $0.$ splits into $00.$ and $10.$ while the branch  $1.$ splits into $01.$ and $11.$  In Figure \eqref{f5}
a few layers of the $p=2$ tree are shown, ending at $u_0 =7$.\footnote{For general $N$ at each node of the tree the incoming branch splits into $N$ branches obtained by writing $0, \ 1,... \ \rm or \it \ N-1$ to the left of the number associated with that node.}

The $p$-adic integers are defined by letting $u_0 \to \infty$, with different choices of branching corresponding to different $p$-adic integers. Choices which terminate, that is whose entries are all zero beyond some value of $u$, are ordinary integers. But whether it is an ordinary integer or not, a $p$-adic integer is best thought of as an entire infinite trajectory on the tree. In other words is is exactly what we earlier called a causal patch. Each causal patch is a point in the ultrametric space of $p$-adic integers.

There is a natural ultrametric distance defined on the $p$-adic integers.  Say that $u$ is the highest power of $p$ that the difference between two $p$-adic integers $x$ and $y$ is divisible by.\footnote{Divisible here means that you can find an $p$-adic integer $z$ such that $x-y=p^{u}z$.}  The $p$-adic distance between $x$ and $y$ is defined as $p^{-u}$.
What this means in terms of the base-$p$ decimal expansions is as follows: starting at the decimal point find the first value of $u$ for which the entries of $x$ and $y$ differ. For example, for $7$-adic integers in base $7$ if
\bea
x \eq ....2363624653412. \cr
y \eq ....1332124653412.
\eea
then the first difference shows up at $u=8$. The $p$-adic distance is defined as $p^{-u}$, which is $7^{-8}$ in this example.  We can define the magnitude of an $p$-adic integer is its distance from zero, that is from
$....0000000.$ Note that in $p$-adic arithmetic the number $1,000,000$ is very small. $1,000,000,000 $ is even smaller.  By inspecting figure \eqref{f5} it is clear that this notion of distance coincides with our previous definition \eqref{u-metric}.

The space $\mathbb{Z}_p$ of $p$-adic integers is a compact space. One indication of the compactness is that the largest distance between $p$-adic integers is $1$.  We have seen that $\mathbb{Z}_p$ is the boundary of the compact version of the cellular model.  For the non-compact case we must generalize to the full $p$-adic number system $Q_p.$  The $p$-adic numbers are defined by allowing a finite number the places to the right of the decimal point to be occupied. For example $....10010.1101$ is a 2-adic number.  Any $p$-adic number $Q$ can be written in the form
\be
x = \sum_{u = -n}^{\infty} x(u)p^u
\ee
with $n$ some finite nonnegative integer.

The space $Q_p$ is non-compact.  We can define the same distance and magnitude as for $\mathbb{Z}_p$, with the only difference being that $u$ can now be negative.  The further the non-zero string extends to the right, the larger the magnitude of $x$ is. $Q_p$ is the appropriate space for the boundary of the non-compact version of the cellular model.  From the cosmological point of view the space $Q_p$ is the future boundary in an analog of the flat slicing of de Sitter space.

\begin{figure}[ht]
\begin{center}
\includegraphics[scale=.35]{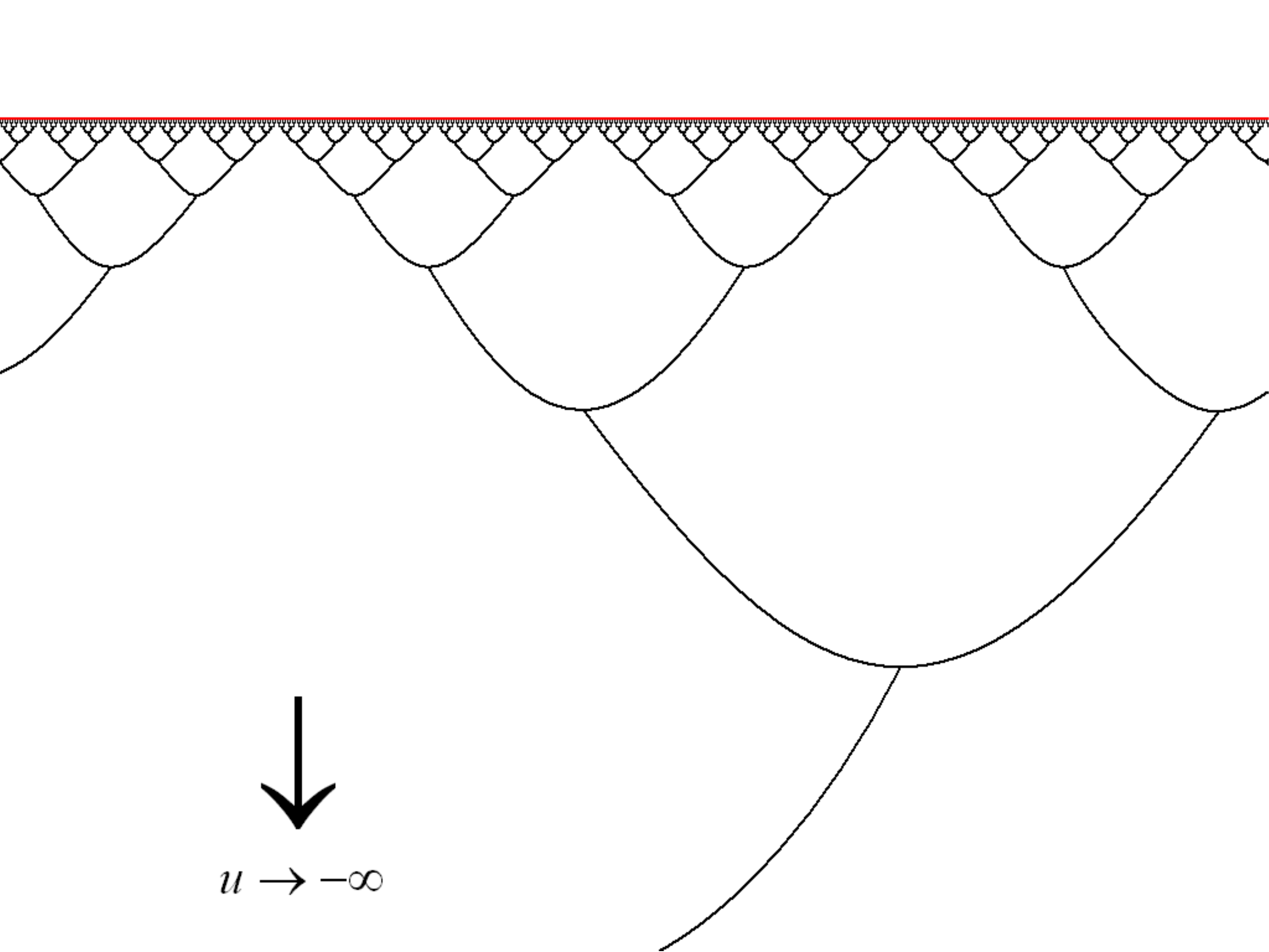}
\caption{The infinite tree for $p=2$ model.  The 2-adic numbers parametrize the red line.}
\label{Qp}
\end{center}
\end{figure}

$Q_p$ can be thought of in terms of trees which extend into the infinite past. Moreover, if an infinite element is added the corresponding tree is identical to a regular tree with every node having $p+1$ ``edges"  coming out of it. Any two nodes are connected by a unique curve formed from the edges,  and such curves can be extended to infinity in both directions to connect pairs of boundary points.  We illustrate this in figure \eqref{Qp}.

The $p$-adic number system is a very rich structure that allows many generalizations from ordinary numbers. The arithmetic processes of addition, subtraction, multiplication and division are all defined. Functions of $p$-adics can be defined. The concept of limits and continuity are simply generalized as is integration. Integration over $p$-adics is most easily explained by first regulating the theory with a maximum value $u_0$ of $u.$ If one has a function on the endpoints---the ``leaves"---of the tree, integration is defined by simply adding the value of the function on every leaf. As in ordinary integration we divide by the number of leaves, $p^{u_0},$ to keep the answer finite,
\be
\int_{p-adic} F(x)dx =\lim_{u_0\to \infty}p^{-u_0}\sum_{leaves} F(leaf).
\ee

\subsection{p-Conformal Symmetry}

The symmetry of de Sitter space is $O(4,1)$ which acts on the bulk geometry as an isometry, i.e., it preserves the distance between any pair of points. The action of the group also induces a group of transformations
on the future boundary which is well known to be  conformal transformations \cite{Witten:2001kn,Strominger:2001pn,Maldacena:2002vr}.  For the most part the conformal symmetry has been studied only in the perturbative theory of de Sitter space, in other words, without bubble nucleation. One of the main lessons of this paper is that the symmetry has significance beyond perturbation theory.    The work of Freivogel and Kleban \cite{Freivogel:2009rf} leads to the same conclusion using a related but distinct framework.

The boundary of de Sitter space
 can be provided with a metric, for example in flat slicing the natural boundary metric is flat.  The $O(4,1)$ conformal transformations of the boundary are not isometries; the distance between two boundary points transforms covariantly.

We will now show that the non-compact version of the cellular model has a remarkable symmetry which parallels the symmetry of de Sitter space. We begin with the case $p=2$
in which the tree splits into two new branches at each node.

So far we have thought of the edges of the tree as being directed. Each edge has a future directed orientation which in our illustrations is upward.
The symmetry that we will explore is not a symmetry of a directed tree: it is a symmetry only if we remove the orientation of each edge. We will see in the next section that for the calculation of correlation functions, the orientation of the tree is irrelevant, as a consequence of detailed balance.

Let us consider the unoriented  tree associated with the 2-adic rationals. That tree is a graph with every node has three edges coming out of it
and there are no closed loops. Such a graph is called a Bethe tree. It is completely homogeneous.  For every pair of nodes, $a$ and $b$  there is a unique path
connecting them. We define the graph-distance $d(a,b)$ between $a$ and $b$ to be the number of edges between the two points.  A transformation of the tree which preserves all graph-distances is an isometry of the tree.

\begin{figure} \begin{center} \includegraphics[scale=.35]{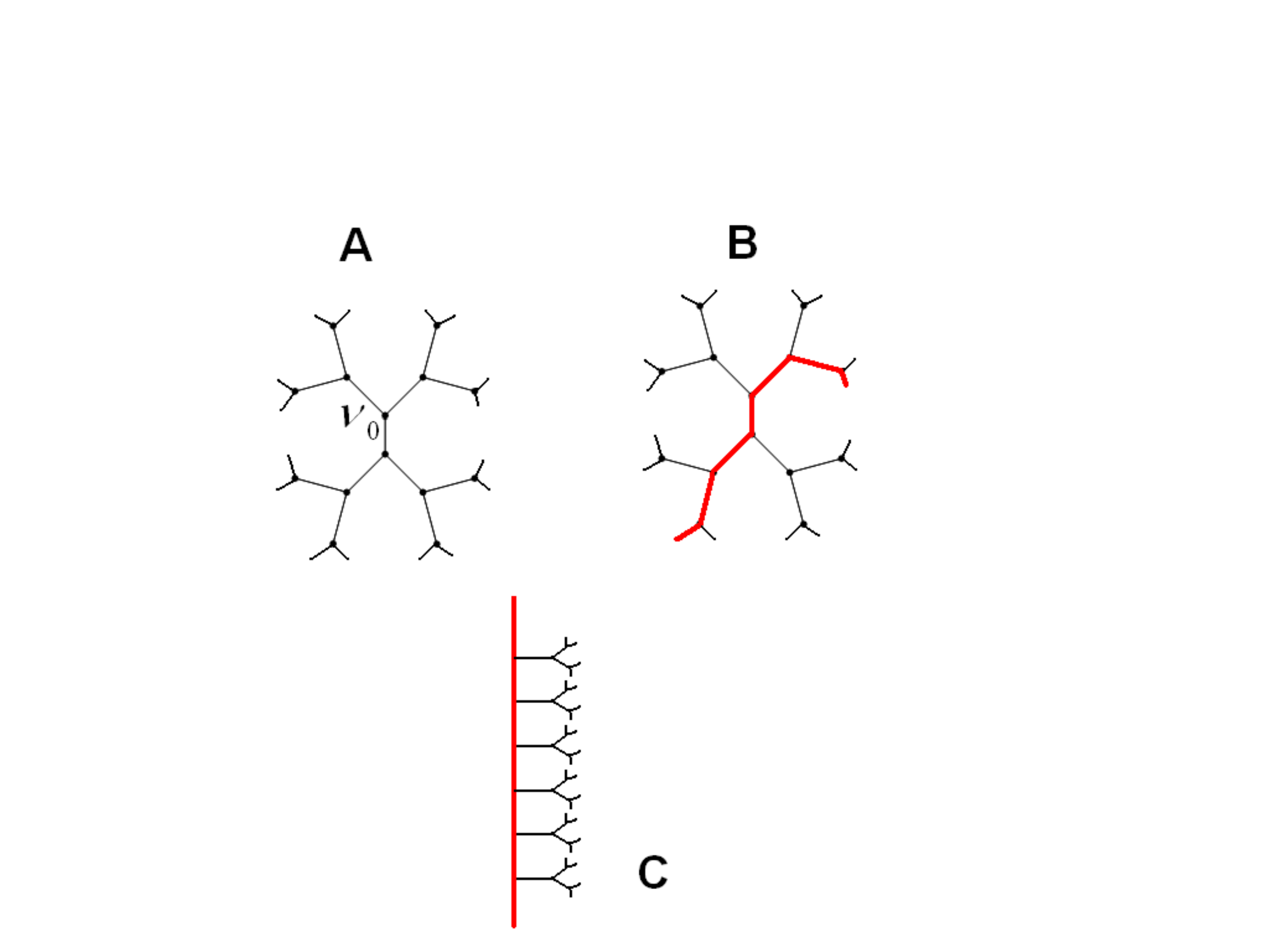} \caption{Bethe Trees. } \label{f8} \end{center} \end{figure}

The  infinite Bethe tree in shown Figure  \eqref{f8} A.
We have arbitrarily located one of the nodes, $\nu_0,$ at the center of the figure. The  symmetry includes an obvious  subgroup, namely,   the permutations among the edges emanating from $\nu_0.$  This symmetry is the analog of the rotation symmetry of the Poincare disc with fixed point at  the origin.

The next symmetry is obtained by considering an infinite trajectory through the tree (as in Figure \eqref{f8} B),
 connecting two  points  on the  boundary of the tree. Think of that trajectory as the trunk of the tree. Define a bush to be a tree that begins with a single link and then branches out as in Figure \eqref{f8}C.
The full tree consists of the trunk with bushes periodically growing out of it.
It is obvious that the tree is symmetric under discrete translation along the trunk. Thus there is a symmetry for which the fixed points are any pair of boundary points. That is the analog of the non-compact transformations that hold two points fixed on the boundary of the disc. When combined with the permutations, a large group is generated.

The general $p$-adic case is similar, the only difference being that a bush is defined to begin with $p-1$ edges.

As in the continuum case, the symmetry transformations induce transformations of the boundary. These transformations act on the space $Q_p.$ The full symmetry group of the tree is quite large; we will be particularly interested in a subgroup $PGL(2, Q_p)$ whose action on the boundary is as the set of fractional linear transformations of a p-adic variable $x$.\footnote{We thank Brian Conrad for a very helpful discussion of these symmetries, in particular for pointing out the relevance of $PGL(2,Q_p)$ as opposed to $SL(2,Q_p)$.}  We can parametrize this as
\be
x^{\prime}={\alpha x + \beta \over \gamma x+ \delta} \qquad \alpha,\beta,\gamma,\delta \in Q_p,
\label{linear fract}
\ee
which is obviously projective.  It is not to hard to see how these transformations can be extended to isometries of the tree. Begin by observing that any pair of boundary points ($p$-adic numbers) $x_a$ and $y_a$
define a node $a$ of the tree. The node is the point of intersection of the causal pasts of $x_a$ and $y_a$.  Now consider a second node, $b,$ defined by boundary points $x_b$ and $y_b.$ The graph-distance between $a$ and $b$
 can be written in terms of the $x.$ The result is
\be
p^{-d(a,b)} = {|x_a-y_a|_p \ |x_ b-y_b|_p \over |x_a-y_b|_p \ |x_b-y_a|_p}.
\label{cross ration}
\ee
This expression is the norm of the $p$-adic cross ratio of the four $x$ values.  It is a familiar fact in ordinary arithmetic that cross ratios are invariant under fractional linear transformations of the form \eqref{linear fract}. The same is true for the $p$-adic numbers. This proves that graph-distances on the tree are invariant under the
$PGL(2, Q_p)$ group.

By contrast, the ultrametric distance between boundary points is not invariant under $PGL(2, Q_p)$ group. Since the ultrametric distance between boundary points is expressed in terms of the time $u$ of their common ancestor, $u$ is also not invariant. The time $u$ transforms covariantly and not invariantly under $PGL(2, Q_p).$ One may wonder whether this fact creates ambiguities when light-cone time is used as a cutoff for purposes of defining a measure.
We will see in section 6 that no such ambiguity occurs.

The action of $PGL(2,Q_p)$ is also \textit{transitive}, meaning that it can send any point on the tree to any other.  In boundary language we can see this by observing that we can send any particular pair of p-adic numbers to any other by acting with an element of $PGL(2,Q_p)$.  Thus this subgroup of the full symmetry is already enough to describe the homogeneity of the tree.

In the next section we will compute correlation functions in the cellular model.  We will find that the asymptotic correlation functions on the future boundary transform covariantly under this $p$-conformal symmetry in a way that closely parallels the conformal symmetry of correlators on the future boundary of de Sitter space. We will also find that the existence of terminals breaks that symmetry in a way that is reminiscent of the persistence of memory effect found in \cite{Garriga:2006hw}.

\subsection{Boundary Volume and the Meaning of the Clock}
\label{clocksect}
Consider a subset of the compact future boundary with regulator time $u_0.$ The volume of the entire future boundary will be normalized to unity. The total number of points is $p^{u_0}$ so each point can be considered to occupy volume $p^{-u_0}$.  Now consider any subset of the boundary, the number of whose points grow as $$p^{u_0}V.$$ Then the volume of the set is defined to be $V.$ Note that since the whole volume is normalized to unity, $V\leq 1.$ We may also write this as
 \be
 \int_{Z_p} 1 dx =1
 \ee
where the integral is over the $p$-adic integers.

The cellular model makes use of a discrete time variable $u$ which activates a p-folding of the number of causal patches at every integer tick of the clock. There are many time variables that are made use of in eternal inflation. A sample is proper time, scale factor time, and light-cone time. At first sight it might seem that $u$ should be identified with scale-factor time. However that is incorrect.\footnote{We thank B. Freivogel for explaining this point to us.}  Consider a node $a$ at time $u_a.$ The causal future of the node defines a set of points on the future boundary. The volume of this set is is related to the time $u_a$ by
\be
u_a = - \log_p V_a.
\label{lightcone time}
\ee
Apart from the fact that the logarithm is base-$p$, this formula is precisely the tree analogue of the definition of light-cone time \cite{Bousso:2009dm} introduced by Bousso, Freivogel, Leichenauer, and Rosenhaus in \cite{Bousso:2010id}.

The authors of \cite{Bousso:2010id} give two definitions of the volume $V_a$ in the continuum theory, but as they argue, the difference is unimportant for many questions. Let's quickly review one of the definitions:  $V_a$ can be defined in terms of a congruence of timelike geodesics constructed to be
orthogonal to some initial surface.  In  \cite{Bousso:2010id} $V_a$ is defined to be
the volume on the initial surface of the subset of the geodesics which enter the causal future of $a.$  More details are provided in appendix A.

This version of light-cone time has an analogue in tree language. For simplicity consider the compact version which begins with a single node representing the initial condition. Imagine that the initial edge contains $p^{u_0}$ geodesics all bundled together (as usual $u_0$ is the cutoff). Every time the geodesics come to a node they evenly divide so that eventually, there is one geodesic per node at the cutoff. We can obviously identify $V_a$ with the number of geodesics entering the causal future of $a.$  If we prefer to use the non-compact version of the model we can define $V_a$ as the fraction of geodesics entering the causal future of $a.$

We can see the analogy of the cellular time to light-cone time in another way. It is evident that as we move up the tree, the number of cells grows as a pure exponential.\footnote{When terminals are present this is not strictly true, since the pruned branches in figure \eqref{f6} no longer grow exponentially.  Our argument here is simplest applied to a landscape with no terminals.}  As we have seen, two distinct nodes at the same level of the tree are unable to communicate in the future.  In de Sitter space this is true for any two points that are not in the same horizon volume, so the natural analogues of the cells in the cellular model are horizon volumes.\footnote{Another way to see this is to observe that, in the model, a single nucleation means a single color change. As we review in appendix A, a bubble nucleation effectively replaces a single horizon volume in the ancestor at the time of nucleation, thus the cells should be horizon size.}  Light-cone time can be usefully characterized as the time for which the number of horizon volumes exponentially grows.  More precisely, under a set of assumptions which we review in Appendix A, the total number of horizon volumes of vacuum $m$ in the stochastic picture of eternal inflation obeys a continuous-time rate equation \eqref{apprate} under which the total number of horizon volumes increases as a pure exponential \cite{Bousso:2009mw}.  We will from here on sometimes refer to the $u$ of the cellular model as light-cone time.

\section{Correlation Functions}

\subsection{The Two-point Function}

In this section we will compute correlation functions evaluated in the equilibrium state of the cellular model. Write the rate equation \eqref{rate equation in comp form} in the form
\be
 P_m(u+1) = G_{mn}P_n(u).
  \label{rate eq in G form}
\ee
If the rate equation satisfies detailed balance, which we will assume for the rest of the section, we can write
\be
G_{mn} = e^{S_m/2}S_{mn}e^{-S_n/2}
\ee
with $S_{mn}$ a symmetric matrix. We can diagonalize $S$ by a basis of orthonormal eigenvectors $(I)_n$, where $I$ labels the eigenvector and $n$ labels the color component. The associated eigenvalue is $\lambda_I$, which we will also write as $p^{-\Delta_I}$
\be
S_{mn} \ (I)_n = \lambda_I \ (I)_m = p^{-\Delta_I} \ (I)_m.
\label{lambda}
\ee

To make the computation simple, it is convenient to define a quantity that we will call a propagator\footnote{The propagator defined here is not the analog of the Feynman propagator which is non-vanishing outside the light cone. Its closest field theory analog is the causal commutator.}. Suppose a cell is known to have color $r$ at time $u_i.$ The propagator is the probability that a descendant of that cell will have color $n$ at time $u_i+u.$  It is given by the $u$-th power of the matrix $G$:
\be
P_{nr}(u) \equiv (G^u)_{nr}= (I)_n \ \lambda_I^u \ (I)_r \ e^{\frac{S_n-S_r}{2}}.
\label{connector}
\ee

Now let us consider the colors on the cutoff surface at $u_0.$ The correlation function $C_{n_a n_b}(a,b)$ is defined as follows: Let
 $a$ and $b$ be two cells at time $u_0.$ $C_{n_a n_b}(a,b)$ is the joint probability that cell $a$ has color $n_a$ and cell $b$ has color $n_b.$

 To compute $C_{n_a n_b}(a,b)$ we begin by tracing back along the causal pasts of both points until we reach the common ancestor cell $r$. Let us suppose that this occurs at time $u_i$ and that the color of the ancestor cell is $n_r$. Then the probability for the colors at $a$ and $b$ to be $n_a$ and $n_b$ is
$$
P_{n_a n_r}(u_0 -u_i)P_{n_b n_r}(u_0 -u_i).
$$
But to compute the correlation function in the equilibrium state we need to multiply this by the fixed point probability $P^{\{0\}}_{n_r} = \frac{1}{\mathcal{N}}e^{S_{n_r}}$ that the ancestor cell had color $n_r$ and then sum over $n_r.$  The normalization factor is $\mathcal{N}\equiv \sum_m e^{S_m}$.
\be
C_{n_a n_b}(a,b) = \frac{1}{\mathcal{N}}\sum_{n_r} e^{S_{n_r}} \ P_{n_a n_r}(u_0 -u_i)P_{n_b n_r}(u_0 -u_i).
\label{CAB}
\ee
Now using \eqref{connector} for the propagators, we get
\bea
C_{n_a n_b}(a,b) \eq \frac{1}{\mathcal{N}}\sum_{n_r,I,J}          e^{S_{n_r}} \ (I)_{n_a} \ \lambda_I^{u_0-u_i} \ (I)_{n_r} \ e^{S_{n_a} -S_{n_r} \over 2} \ (J)_{n_b} \ \lambda_J^{u_0-u_i} \ (J)_{n_r} \ e^{S_{n_b} -S_{n_r} \over 2} \cr \cr
\eq \frac{1}{\mathcal{N}}\sum_{I}
\lambda_I^{2 (u_0-u_i)} \ (I)_{n_a} \ (I)_{n_b} \ e^{S_{n_a}+S_{n_b} \over 2}.
\eea
In the last step, we used the orthonormality of the eigenvectors to set $I=J$. One can also write the correlation function in the $I,J$ basis in which it is diagonal and extremely simple,
\be
C_{IJ}(a,b) \equiv \mathcal{N} \sum_{n_a n_b}\ C_{n_a n_b}(a,b) \ (I)_{n_a} \ (J)_{n_b} \ e^{-\frac{S_{n_a}+S_{n_b}}{2}} =  \delta_{IJ}
 \lambda_I^{2(u_0-u_i)}.
\ee
As a final step, we would like to extrapolate this correlator to the future boundary. This can be done by pushing the points $a,b$ up the tree, and stripping off vanishing factors. Mathematically, we let the level $u_0$ run to infinity, and define sequences $\{a_{u_0}\},\{b_{u_0}\}$ in the causal futures of $a$ and $b$ respectively. These sequences define $p$-adic numbers $x$ and $y$. To keep the result finite, we will multiply by a wave function renormalization factor of $\lambda_I^{-u_0}$ for each external leg. The result is
\be
\langle \mathcal{O}_I(x)\mathcal{O}_J(y)\rangle \equiv \lim_{u_0\rightarrow\infty} C_{IJ}(a_{u_0},b_{u_0})\lambda_I^{-u_0}\lambda_J^{-u_0} = \delta_{IJ}\lambda_I^{-2u_i}.
\ee
Recalling that $\lambda_I=p^{-\Delta_I}$ and recognizing $p^{-u_i}$ as the $p$-adic distance $|x-y|_p$, we have
\be
\langle \mathcal{O}_I(x)\mathcal{O}_J(y)\rangle = \frac{\delta_{IJ}}{|x-y|_p^{\,2\Delta_I}}.
\ee

The correlator has the precise form of a two-point function of primary operators
in a conformal field theory. The fact that the correlation function has an overall
scaling behavior is not surprising. It follows from the scale invariance of the fixed point. The unexpected property is the diagonal nature of $\langle \mathcal{O}_I(x)\mathcal{O}_J(y)\rangle.$ In a conformal field theory this property is a consequence of invariance under the special conformal transformations which have $x$ and $y$ as fixed points. In the cellular model it is a consequence of the orthogonality of the eigenvectors which, in turn, followed from detailed balance.

\subsection{The Three-point Function}
The agreement of the 2-point function with the rules of conformal field theory is not a fluke. Let us move on to the 3-point function. Let the three points be denoted $a,$
$b,$ and $c,$ and let them have colors $n_a,$ $n_b,$ and $n_c.$  Recalling that triangles in ultrametric geometry are always tall isosceles, we take the short edge to be between $a$ and $b$. Tracing back to their common ancestor cell $d$, we label that cell with color $n_d$. $a,$$b,$ and $c$ are located on the cutoff surface $u_0$ and $d$ is at time $u_d.$ The common ancestor of all three points is called $e.$

Finally, define
\bea
u_0 -u_d &\equiv& u \cr
u_d -u_e &\equiv& v.
\eea

The 3-point function is given by
\be
C_{n_a n_b n_c}(a,b,c)= \frac{1}{\mathcal{N}}\sum_{n_d,n_e} e^{S_{n_e}}P_{n_d n_e}(v)P_{n_c n_e}(v+u)P_{n_a n_d}(u)P_{n_b n_d}(u).
\ee
To compute this correlation function we insert \eqref{connector} and use the orthonormality  of the eigenvectors. As above, that the answer simplifies somewhat in the eigenvector basis. After a straightforward computation, one finds
\begin{align}
  C_{IJK}(a,b,c)& \equiv \mathcal{N}^{3/2}\sum_{n_a n_b n_c} C_{n_a n_b n_c}(a,b,c) \ (I)_{n_a} \ (J)_{n_b} \ (K)_{n_c} \ e^{-{S_{n_a}+S_{n_b}+S_{n_c} \over 2}} \notag \\
  &=\mathcal{N}^{1/2}\sum_n e^{-S_n\over 2} \ (I)_n \ (J)_n \ (K)_n \ p^{-(\Delta_I + \Delta_J -\Delta_K )u -2\Delta_K(u+v)}.
\end{align}

  The factor $p^{-(\Delta_I + \Delta_J -\Delta_K )u -2\Delta_K(u+v)}$ may look unfamiliar, but after stripping off external leg factors as for the two point function, and taking the limit to the future boundary, the result can be expressed in terms of the $p$-adic distances in a completely symmetric way.

  \begin{align}
  \langle \mathcal{O}_I(x)\mathcal{O}_J(y)\mathcal{O}_K(z)\rangle & \equiv \lim_{u_0\rightarrow\infty} C_{IJK}(a_{u_0},b_{u_0},c_{u_0})\lambda_I^{-u_0}\lambda_J^{-u_0}\lambda_K^{-u_0}  \\
  & =\frac{C_{IJK}}{|x-y|_p^{\,\Delta_I+\Delta_J - \Delta_K} |x-z|_p^{\,\Delta_I+\Delta_K-\Delta_J} |z-y|_p^{\,\Delta_K + \Delta_J - \Delta_I}}.
  \end{align}
  Here the structure constants are
  \be
  C_{IJK} = \mathcal{N}^{1/2}\sum_n e^{-S_n \over 2} \ (I)_n \ (J)_n \ (K)_n.
  \ee
  This three-point function has the exact form of a correlation function of three scalar operators of dimension $\Delta_{I,J,K}$ in a conformal field theory, except expressed in terms of $p$-adic distance \eqref{u-metric}.

The $C_{IJK}$ are the analog of the operator product coefficients.  They obey the basic associativity relation

\be \label{assoc}
\sum_H C_{IJH}C_{HKL} = \sum_H C_{IKH}C_{HJL},
\ee
guaranteed by orthonormality of the eigenvectors.

\subsection{General correlators and PGL(2,Qp)}
\begin{figure}[ht]
\begin{center}
\includegraphics{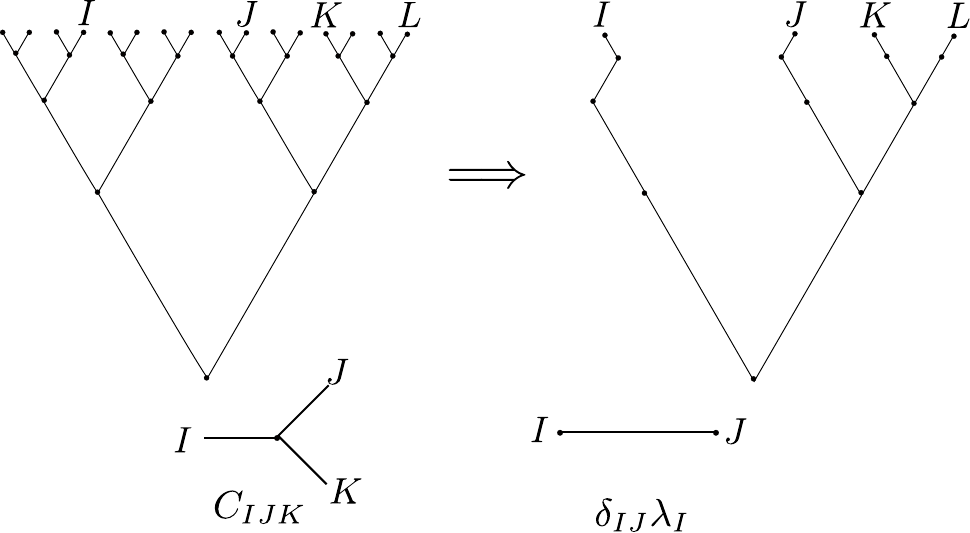}
\caption{Masking unnecessary links (top) gives the diagram from which the correlator is computed using graphical rules (bottom). The result for the example shown is $\sum_H C_{HIJ}C_{HKL}\lambda_I^5\lambda_J^3\lambda_H\lambda_K^2\lambda_L^2$.}
\label{rules}
\end{center}
\end{figure}

  A computation similar to the above can be done for higher point functions. The result in the eigenvector basis can be compactly summarized for any correlation function. Begin with the portion of the tree containing all points in the correlator and their common ancestors. Mask all links that aren't necessary to connect the correlator insertions (as shown in Fig \eqref{rules}). Now apply a one-step propagator to each unmasked link, and assign structure constants to the vertices.

These rules uniquely determine the correlation functions at finite points on the tree. To compute boundary correlators, we take points to infinity, where they define $p$-adic numbers. Stripping off factors of $\lambda_I^{u_0}$ makes the resulting expression a function of the $p$-adic distances between the points.

To show that the conformal properties of the two- and three-point functions persist to higher multiplicity, we will show that the extrapolated functions are covariant under $PGL(2,Q_p)$ transformations. $PGL(2,Q_p)$ acts on the tree as a group of isometries, leaving the number of links between points unchanged. Since the graphical rules depend only on these distances, the correlation functions at finite points on the tree are invariant under the group action. The extrapolated correlators are not invariant, though, since the level $u_0$ transforms under $PGL(2,Q_p)$, and consequently the stripping factors $p^{-\Delta_I u_0}$ change.

Let's evaluate this change. Any vertex in the tree can be defined as the most recent common ancestor of two points $x,y$ on the boundary, with the level $u$ given by $p^{-u} = |x-y|_p$. The action of $PGL(2,Q_p)$ on the boundary now conveniently defines the action on the tree. Let $\{y_k\}$ be a sequence of $p$-adics converging appropriately to $x$. Then $\{(x,y_k)\}$ define a sequence of tree vertices with increasing level, approaching the future boundary at point $x$. We can use this sequence to evaluate the asymptotic ratio of $p^{-u}$ before and after the $PGL(2,Q_p)$ transformation $f$ as a limit
\be
\lim_{k\rightarrow \infty} \frac{|f(x) - f(y_k)|_p}{|x-y_k|_p} = |f'(x)|_p.
\ee
As a result, the new stripping factor for $\mathcal{O}_I(x)$ after the $PGL(2,Q_p)$ transformation is $|f'(x)|^{\Delta_I}$ times the old one: extrapolated operators transform covariantly, in a manner precisely analogous to conformal primaries in CFT.

Before moving on, we will make a few comments.

\begin{itemize}
\item Detailed balance was crucial in establishing the conformal properties of the correlators. Without it, the graphical rules depend on orientation of links. Since orientations are not preserved by the action of $PGL(2,Q_p)$ on the tree, the resulting correlators aren't invariant.

\item The rules for building correlation functions have an equivalent OPE description
\be
\mathcal{O}_I(x)\mathcal{O}_J(y) \rightarrow \sum_K \mathcal{O}_K(y) C_{KIJ} |x-y|_p^{\,\Delta_K-\Delta_I-\Delta_J},
\ee applicable whenever $y$ is the closest insertion to $x$ and vice versa. $\mathcal{O}_0$ acts as the identity in this operator algebra.  As usual, factorization on $\mathcal{O}_0$ gives clustering.
\item The four-point function can be written in terms of $\Delta = \sum_i \Delta_i$ and the cross-ratio $x = \frac{x_{12} x_{34}}{x_{14} x_{23}}$ as
\be
\langle \mathcal{O}_1(x_1)\mathcal{O}_2(x_2)\mathcal{O}_3(x_3)\mathcal{O}_4(x_4)\rangle = \prod_{i<j}|x_{ij}|_p^{\,\Delta/3 - \Delta_i - \Delta_j}\sum_{H}C_{H12}C_{H34}|x|_p^{\,\Delta_H - \Delta/3}
\ee
where we've assumed that $x_1$ is the closest insertion to $x_2$ and vice versa. When $x_1$ is equidistant from $x_2$ and $x_3$ there are two different ways to evaluate this correlator.   That they agree follows from the associativity of structure constants (\ref{assoc}).

 If $p>2$, it is possible for the masked tree to have vertices with, say, four lines coming out of them. These are treated as coincidence limits of two three-way vertices, with a zero length propagator in between. Of course, there are multiple ways to contract the indices in this limit. Equation (\ref{assoc}) ensures that they all agree.

\item The correlation functions become very non-gaussian if the differences in entropies are large, in the sense that some $C_{IJK}$'s are parametrically larger than 1.

\item The model can alternatively be formulated as a generalized Ising model on the tree, where the spin variable runs over colors. The Boltzmann weights on the links are determined by $S_{nm}$, and those on the nodes by $Z_{nn}$.

This is a generalization of the free scalar field construction in \cite{Zabrodin:1989cd} and the Ising model calculations in \cite{1991CMaPh.139Q.433Z}.
\end{itemize}

\section{Time's Arrow, Terminals, and Fractal Flows}

The arrow-of-time question---Why does time have a direction?--- is one of the grand  questions of  cosmology.  It is also the problem of why the history of the universe is not a Boltzmann fluctuation\footnote{A colorful name for the same difficulty is the problem of Boltzmann Brains. But even if one believed that life requires a universe more or less similar to our own, the problem of Boltzmann fluctuations is still a paradox.}. It troubled Boltzmann himself, and in its modern de Sitter  version it was raised  in \cite{Dyson:2002pf}.  In a thermodynamically closed universe (the causal patch of de Sitter space is thermodynamically closed)  there are just too many ways to make a universe in which life can exist---too many trajectories in phase space---and if all micro-states are equally likely, the probability that world resembles ours is exponentially negligible.

The arrow-of-time problem is often stated as a problem of initial conditions: why did the universe begin in a state of very low entropy?\footnote{This formulation is usually attributed to Penrose.  Other more recent discussions include \cite{Albrecht:2002uz,Carroll:2004pn}.} However, there is more to it than that.  If a closed system starts in a state of low entropy, there will be a transient behavior, with a time-arrow, during which conventional cosmology may occur before thermal equilibrium is achieved. However, after thermal equilibrium,  there will  be an infinite future without a time-arrow, in which repeated Boltzmann fluctuations and Poincare recurrences produce an endless sequence of inhabited worlds, the overwhelming majority of which are inconsistent with observation. Eternal inflation without terminals cannot solve this problem.

Consider the description along a single causal patch. According to our assumptions thus far, if there are no terminals, the causal patch tends to an attractor with uniform probability for every microstate. Detailed balance insures that within such a patch there will be no time-arrow. Indeed, it is exactly the lack of a time-arrow that makes the attractor conformally invariant in the bigger global picture. A conformally invariant fixed point is not a good feature; it is a disease.

It has been understood for a while that the existence of terminal vacua changes the story in a very positive direction\cite{Linde:2006nw,DeSimone:2008if,Bousso:2008hz}, and the sharpest statement of these ideas is in a recent paper by Bousso \cite{Bousso:2011aa}.  The relevant mechanism is simply illustrated in the cellular model.  The conformally invariant fixed point in the theory without terminals is replaced by a new kind of attractor that does have a time-arrow.

In the presence of terminal vacua, the co-moving coordinate volume  eventually becomes solidly dominated by terminals. Eternal inflation does not end, but it is restricted to a fractal of coordinate volume  that decreases like  $\lambda_D^u$   ( $\lambda_D$ being the dominant eigenvalue of the rate equation \ref{nonterminal rate equation}). Despite the fact that the coordinate volume goes to zero, it is still possible to define a collection of non-trivial \it conditional correlation functions \rm on future infinity:
\bea
&&C_{n_a}(a) \cr  \cr
&& C_{n_an_b}(a,b) \cr \cr
&& C_{n_an_bn_c}(a,b,c) \cr \cr
&&............
\eea

The  one-point function $C_{n_a}(a)$ is defined to be the probability that the vacuum at point $a$ has color $n_a,$  given that $a$ has not been swallowed by a terminal bubble. It can be identified with the dominant eigenvector $P^{\{D\}}_m.$
Similarly $C_{n_an_b}(a,b)$ is the joint probability that $a$ and $b$ have color $n_a, \ n_b,$
given that neither point has been swallowed by terminals, and so on.
These conditional correlation functions define a new kind of non-conformal attractor that is not a conventional field theory. We will explore its properties.

 Let's return to the two point function. Equations \eqref{connector} and \eqref{CAB} are unchanged by the effects of terminals with one exception. The factor $P^0
_n =\frac{e^{S_n}}{\mathcal{N}}$ represented the probability that the color of the common ancestor is $n$ in the equilibrium state. When there are terminals the equilibrium state should be replaced by the dominant eigenvector with non-normalized probabilities\footnote{Since the sum of these probabilities is not conserved there is no natural normalization, any particular choice just sets an initial time where volume started leaking into terminals.}
\be
P^{\{D\}}_n = e^{S_n \over 2}  \ (D)_n \  p^{-\Delta_D u_i}
\ee
Thus

\be
C_{n_a n_b}(a,b) = \sum_{n_r}
p^{-\Delta_D u_i} e^{S_{n_r} \over 2} \ (D)_{n_r}
P_{n_a n_r}(u_0 -u_i)P_{n_b n_r}(u_0 -u_i),
\label{CABD}
\ee
which can be written,
\be
C_{n_a n_b}(a,b)= \sum_{I,J}\left\{   \sum_n e^{-S_n \over 2} \  (D)_n \  (I)_n \
(J)_n
 \ p^{(\Delta_J+\Delta_I-\Delta_D ) u_i}   \right\}
\left[
(I)_{n_a} \
(J)_{n_b}  \ e^{S_{n_a} +S_{n_ b} \over 2}  \  \ p^{-(\Delta_I+\Delta_J)u_0} \right].
\ee

The factors in the square brackets are removed by going to the eigenvector basis and extrapolating to the boundary, which gives
\be
\langle \mathcal{O}_I(x)\mathcal{O}_J(y)\rangle=   \sum_n e^{-S_n \over 2} \  (D)_n  \ (I)_n  \ (J)_n
\left({1 \over |x -y|_p}  \right)^{(\Delta_J+\Delta_I-\Delta_D ) }.
\label{insertion}
\ee

This formula has an interesting structure. $\langle \mathcal{O}_I(x)\mathcal{O}_J(y)\rangle$ has the form of a limit of a three point function in which one of the points is at $p$-adic infinity.  In other words the effect of terminals resembles the insertion of an additional operator with dimension $\Delta_D.$  This insertion breaks the $p$-conformal symmetry.  It is expected that the existence of terminals causes this breaking \cite{Garriga:2006hw}, but that it does so in such a simple way is surprising.  Moreover the higher point correlators can still be computed using an OPE, but in a theory where $\mathcal{O}_D$ has a one-point function and no operator of dimension zero ever appears.  Because of this the correlators do not cluster in the conventional sense.  At long distances between groups of operators correlation functions reduce to correlators of $\mathcal{O}_D$. Therefore they cannot define field theories of any conventional kind.  We will call this unusual structure a \textit{fractal flow}.

The breaking of the conformal symmetry in a fractal flow is a useful feature. Without it the time variable $u$ is not unique; it transforms under $p$-conformal transformations. But once the extra insertion is introduced the only remaining symmetry transforms the surfaces of fixed $u$ into themselves.

The fractal flow depends on the existence of an initial condition but not on its details. The only important feature of the initial condition is that it allows eternal inflation to take place. There are initial conditions that would preclude eternal inflation; for example an initial condition in which all space is occupied by a terminal vacuum. The theory of fractal flows cannot tell us why the
initial condition is not of this type. But what it does imply, is that if eternal inflation occurs, the final attractor will have an arrow of time.

  \setcounter{equation}{0}
\section{Multiverse Fields}

As with all models, there are those things which generalize to the phenomena that we are trying to model, and those that do not. From the former we can learn lessons. The latter can mislead us. Separating the good from the bad always involves a bit of guesswork.

First of all, the cellular model is not an approximation to continuum eternal inflation in the same way that lattice gauge theory is an approximation. It is a distinctly different structure which does not approach the continuum theory in any limit. The situation is more like modeling $3+1$ dimensional gauge theory by lower dimensional versions. Some things will be totally wrong---the propagation of signals for example---but other things like the existence of a confining phase will be right.

One thing which is totally wrong  about the  original lattice-version of  the cellular model  is its notion of spatial proximity. One might expect that if two cells of the lattice are neighboring,  then the correlations between them will be strong. But two cells can be adjacent, and still very distant, in the sense that you have to go a long way back to find a common ancestor cell. Such cells are weakly correlated. On the other hand, the idea that correlations between causally disconnected points are due to the past overlap of their causal pasts is common to both theories.

The thing that we stripped away when we went from lattice to tree version was the concept of space-like separation between  out-of-causal-contact points. From the point of view of continuum de Sitter space it may seem an unreasonable thing to eliminate. The truth however, is that there is no clear meaning to out-of-causal-contact space-like separation  in de Sitter space. In de Sitter space, there are no space-like geodesics between distant points,  which can be used to define space-like geodesic distance. As we will describe below, the best definition of  distance parallels  the definition  ultrametric distance on the tree.

The correlation function for a scalar field $f$ in de Sitter space can be calculated by analytic continuation from the Euclidean sphere. On the sphere it is obvious that the correlation function depends only on the geodesic distance, between the points. Consider that continuation to de Sitter space in flat-slicing coordinates in which the metric has the form
\be
H^2 ds^2 =-d\omega^2 + e^{2\omega}dx^i \ dx^i.
\label{ds metric}
\ee

The two points $a$ and $b$ are spatially separated by a fixed coordinate separation and the time coordinates $\omega$ are asymptotically large. This implies that the points are out of causal contact.

In continuing from the Euclidean sphere we immediately encounter the problem that there is no geodesic connecting such points. However the formula for the geodesic distance between the points can be continued, although it becomes complex. One easily finds  (see appendix B of \cite{Harlow:2010my})  that for large $\omega$ and fixed $x$ the complex proper time between the points, in units of $H^{-1},$  has the form
\be
\tau =  i\pi + \omega_a + \omega_b + 2 \log{|x|}
\label{tau}
\ee
where $|x|$ is the coordinate distance between the points. For a light scalar field the correlation function has the form
\be
\langle f(a) f(b) \rangle \sim e^{-\Delta \tau}
\ee
which gives the usual power law in $|x|.$

But now let us look more closely at the geometric meaning of  \ref{tau}. The two points $a,b$ are out of causal contact, but if we run them backward toward the past at fixed $x$ they will eventually be within distance $H^{-1}.$ In other words they were once within the same causal patch. This occurs where the causal pasts of the two points first intersect in a common ancestor patch.
The time of this  intersection is
\be
\omega_i= -\log{|x|}.
\ee

Now consider the proper time between $a$ and the intersection time. Call it $\tau_a.$
\bea
\tau_a \eq \omega_a  - \omega_i  \cr \cr
\eq  \omega_a  + \log{|x|}.
\label{tau a}
\eea

Similarly for $b,$

\be
\tau_b =  \omega_b  + \log{|x|}.
\label{tau b}
\ee

Thus we see that the real part of  \ref{tau} is just proportional to the sum of the two proper times between the common ancestor causal patch and the points $a,b.$  The imaginary part of \ref{tau} is the space-like distance between the points at the time $\omega_i.$ Evidently one can think of the complex geodesic between $a$ and $b$ as consisting of three segments: two time-like segments from the points to the common ancestor, and a short space-like distance in the ancestor patch.

It is clear from this construction that the only  correlation between $a$ and $b$ is due to the same mechanism as in the cellular model: correlation due to the fact that the points were once in the same causal patch.
In one case the mechanism leads to conventional de Sitter correlators: in the other case the same mechanism leads to ultrametric correlators.

The cellular model is really about phenomena that take place on scales so large that they involve points which are out of causal contact. It is also about a sector of the theory, that from the boundary point of view,  involves fields of extremely small dimension $\Delta \ll 1$. From the bulk point of view these fields only vary between different  Hubble patches. One of the lessons of the cellular model is that the number of such fields is enormous.

To see this, consider the fields that project onto particular colors, whose correlators we computed in the previous section.  There is such a field for each color $n$ in the landscape. The fields associated with the eigenvectors of the rate matrix $(I)$ are equal in number. A brief look at the rate equations shows that the dimensions $\Delta_I$ are of the same order of magnitude as the rates $\gamma.$ Assuming that the vacua are much longer lived than the Hubble times, the dimensions are much less than one. The cellular model describes the sector of these very low dimensional operators, which from the bulk point of view means fields that vary on scales larger than the horizon. These fields could be called multiverse fields.

The multiverse fields are extremely non-gaussian. We can see that from the correlation functions of section 3. If we normalize the two point functions canonically, then large differences in entropies make the three point functions very large. This is not particularly surprising;  bubble nucleation is expected to create non-gaussian effects.

Does the non-gaussian behavior of multiverse fields ever lead to observable effects? The answer may be yes.  If cosmological parameters are in a lucky window, bubble collisions may be visible on the sky. These bubble collisions would have the same kind of non-gaussian behavior as multiverse fields. But probably more interesting is the role of multiverse fields in determining the statistics of vacua: in other words the measure.

 \setcounter{equation}{0}
 \section{The Measure Problem}

 There have been a number of recent studies of the so-called measure problem\footnote{An original reference on the measure problem is \cite{LLM}. For a review with further references, see \cite{guth}.} that
 suggest that some promising progress has been made. Here is a partial list: \cite{DeSimone:2008bq}\cite{Bousso:2008hz}\cite{Bousso:2010id}. Each proposal is based on a preferred time variable which is used to regulate the infinities implicit in an eternally inflating multiverse. For example  \cite{DeSimone:2008bq} uses scale-factor time which is defined in such a way that the volume of a comoving region grows exponentially. Reference \cite{Bousso:2010id} uses light-cone time defined so that the number of causal patches grows exponentially.  The results are not too different and have a common feature. In all these ``geometric" measures, the probability for a given value of the cosmological constant is dominated (for large values) by a factor
\be
\exp{\left(-3H\tau_{obs}\right)}
\label{exp -3Ht}
\ee
 where $\tau_{obs}$ is the ordinary FRW time at which observations are made.\footnote{This is the same factor that leads to the ``Guth-Vanchurin Paradox'' \cite{Guth:2011ie}.}  The implication is that the typical value of $H$ is no larger than $\sim \tau_{obs}^{-1}.$ We may consider this to be an explanation of the so-called coincidence problem.

Different proposals  \cite{DeSimone:2008bq}\cite{Bousso:2008hz}\cite{Bousso:2010id} differ in the details of the measure at small values of $H\tau_{obs}$ where all kinds of details are involved, but they all agree on \eqref{exp -3Ht} for large $H.$
There is one feature of the light-cone proposal which is particularly intriguing, namely, it is precisely equivalent to the causal patch measure which is defined in terms of local observations within a single causal patch \cite{Bousso:2009dm}\cite{Bousso:2009mw}.

In this section we will see that the stripped down version of the cellular model retains exactly the right features to explain \eqref{exp -3Ht} and the local-global duality of \cite{Bousso:2009dm}\cite{Bousso:2009mw}. In what follows we will derive the relative probabilities for vacua of different Hubble rates $H_n$ in the cellular model.

 We will need to incorporate a few additional ingredients into the cellular model. The new ingredients concern the existence of observers, although they do not depend on any detailed assumptions about the nature of life or intelligence.  First, the usual  assumption of typicality:  the probability of an observation of a given type is proportional to the number of such observations made under the cutoff surface.  It is not our purpose to justify this assumption but only to show how the geometry and causal structure of the multiverse influence the answer.

The possibility of observers in terminals is a confusing issue which we will not address in this paper.  The tree model is not well suited to discussing the interior of terminals, and in more detailed work on the issue \cite{Linde:2006nw,DeSimone:2008bq,Bousso:2008hz,Bousso:2010id} their importance has depended on the detailed choice of measure.  For simplicity we will just assume that observers do not exist in terminals and accordingly we prune the tree.  

Next, we need to define the concept of  nucleation of a bubble of a given vacuum type $n.$
 Consider a node of type $n$ somewhere on the tree.  The causal past is defined by tracing back along a unique series of edges of the tree. 
 Eventually as one works backward, a point will occur where the color is no longer $n.$
 That point is the nucleation point $a_n$. If all rates $\gamma$ are very small, then what grows out of the nucleation point will be mostly of type $n.$

Another  standard assumption is  that observers  in the type $n$ environment can exist  for a limited range of proper time, subsequent to the nucleation event. For simplicity we take that period to be concentrated around a proper time  $\tau_{obs}.$   This proper time can be translated to a number of links. In the $n$-vacuum the proper time of a segment composed of $L$ links is
 \be
 \tau = {L \over H_n}.
 \ee
Setting this equal to $\tau_{obs}$ gives the number of links $u_{obs}$ from the nucleation point to the place where observations take place,
\be
u_{obs} = H_n \tau_{obs}.
\ee

Let us also assume that when the nucleation of a bubble of type $n$ takes place, or shortly thereafter, an amount of matter is created that will eventually be assembled into a number of observers $\nu_n.$
It is generally a function of the vacuum type, but we make the reasonable assumption that it is independent of $H_n$ and of $\tau_{obs}$ for $H_n\tau_{obs}\sim 1$, which we will see below is where the measure is concentrated.


Finally, to regulate the inevitable infinities of eternal inflation we introduce a cutoff time $u_0.$ However, the $p$-conformal symmetry raises a question: since $u$ transforms non-trivially, might there be a danger that the measure will depend on a choice of frame? Let us proceed, ignoring the question for the moment.

In principle we would like to count all observations that occur below the cutoff surface, but because populations exponentially increase, in an eternally inflating world it is sufficient to count observations that take place at the cutoff. Thus one defines the  the measure ${\cal{M}}(n)$ to be proportional to number of observations occurring at time $u_0$, in bubbles of type $n,$ anywhere in the multiverse. At the end of the calculation, $u_0$ is allowed to tend to infinity.  This choice of cutoff is natural in the model and in particular respects the symmetries but it is not the only possible choice.

There are several factors that go into  ${\cal{M}}(n)$. The first and most obvious is $\nu_n,$ the number of observers that form in a  single bubble of type $n.$

The next relevant consideration is the color $m$ of the immediate ancestor of the bubble. If the ancestor color is assumed to be $m$ then we must include a factor of the probability for color $m$ in the statistical ensemble. Assuming the statistical ensemble is given by the dominant eigenvector gives a factor $P^D_m.$ In addition we must include the probability $\gamma_{nm}$ that the vacuum $m$ decays to $n.$
Summing over the possible ancestor vacua leads to the factor
$$\sum_m \nu^n \ P^D_m \ \gamma_{nm}.$$

The final and most important factor comes from the fact that if the observations take place at time $u_0,$ the bubbles must nucleate at time $u_0 - u_{obs}.$
Consider the total number of sites available on the tree at time $u_0.$
That number is
$8^{(u_0 - u_{obs})}.$  With this factor included, the measure becomes proportional to
\be
   {\cal{M}}(n) \sim     \nu_n  \  \sum_m \ \gamma_{nm} \ P^D_m  \  \ 8^{(u_0 - H_n \tau_{obs})}
\ee

Now we see the possible danger: the answer appears to depend on the cutoff surface $u_0$ through the factor
$8^{u_0}$. But this factor represents the  exponential growth of all populations and should be factored out of  relative probabilities. The relative probabilities are not only independent of the value of $u_0.$  All dependence on the specific choice of conformal frame has canceled out. Thus there is no ambiguity due to the fact that $u$ transforms covariantly.

After dropping the cutoff dependent factor the remaining measure is
\be
   {\cal{M}}(n) \sim     \nu_n  \  \sum_m \ \gamma_{nm} \ P^D_m  \  \ 8^{- H_n \tau_{obs}}.
\ee
The factor that interests us most is the
 one containing the hubble constant $H_n,$
\be
{\cal{M}}(n) \sim 8^{ - H_n \tau_{obs}}
\label{cc measure}
\ee
In the literature on the subject it is generally assumed that the remaining factors are not statistically correlated with $H_n.$ In that case \eqref{cc measure} provides  a measure for the cosmological constant. As already emphasized in earlier papers the formula in \eqref{cc measure} leads to a highly successful correlation between the value of the cosmological constant and the time of observation.

Finally, we note that the local-global duality of \cite{Bousso:2009dm} \cite{Bousso:2009mw} is also manifested in the  cellular model. Consider following a causal patch from its beginning until it either reaches the boundary of the tree or runs into a terminal vacuum. Along the way the patch will make transitions to different colors and the question is how many observers are encountered in each vacuum type. In this form, the question does not depend on the existence of a cutoff $u_0.$ Let us suppose the causal patch makes a transition to vacuum type $n$ at time
$u.$ The material for the  $ \nu_n $ observers that is created shortly after the nucleation will be distributed among the branches that grow out of the nucleation point. By the time the observations take place the number of branches has grown from one to $8^{u_{obs}}= 8^{H_n \tau_{obs}}.$ Thus the relative number of observations that takes place in the causal patch is given by \eqref{cc measure}.

The local-global duality of \cite{Bousso:2009dm} \cite{Bousso:2009mw} is an attractive feature of the light-cone measure that is not shared by other measure proposals.

 \setcounter{equation}{0}
 \section{Conclusion}
de Sitter space, and the cellular model without terminals, are two realizations of a common set of definitions and postulates. The postulates begin with the existence of a transitive relation which allows us to define the concepts of causal future and past; causal diamonds; and causal contact or the lack of it; and eventually causal patches. To these we added the postulates: world-lines eventually fall out of causal contact: spacetime is causally and metrically homogeneous. (An interesting question is whether there are any other realizations, apart from trivial generalizations such as products of de Sitter space and homogeneous spaces?)  Both structures possess powerful symmetries---conformal symmetry in the de Sitter case: $p$-conformal symmetry in the cellular model---which control the behavior of correlation functions.

Generic initial conditions break this symmetry, both in the model and in de Sitter space. In the model without terminals, we found that detailed balance is enough to restore the symmetry to the late-time correlation functions. We suspect that a similar condition holds in de Sitter space, since detailed balance is required for the droplet-within-droplet symmetry of field configurations demanded by conformal inversions (see section 7 of \cite{Freivogel:2006xu}).

Terminal vacua, which play a very important role in the theory of eternal inflation, can be added to the cellular model, resulting in the breaking of conformal or $p$-conformal symmetry. The model is analytically tractable and we find a very simple effect of the terminals; the correlators still factorize with a local OPE but the conformal symmetry is broken and the identity is removed from the operator algebra.  We call this structure a fractal flow.  No analogous analysis has been done in the continuum theory, but given the similarity of the two we are led to conjecture that the effect of terminals on correlators is the same.

The cellular model concentrates on the large scale behavior of multiverse fields as defined in section 4. The behavior on scales smaller than a causal patch is lumped up into the $e^S$ microstates in each vacuum of the landscape. These multiverse fields define a sector comprising a huge discretuum of extremely low dimension operators with highly non-gaussian correlations. The dimensions are
 determined by the eigenvalues of rate equations that parallel the equations of the continuum theory, with the exception that in the cellular model the equations are rigorous. The multiverse fields describe the statistics of how the landscape is populated.

The time variable of the cellular model is the exact analog of the light-cone time in the continuum theory, and provides a natural cutoff. By adding a few additional bits to the model it can be applied to the measure problem. We find that the resulting measure is essentially the light-cone measure of \cite{Bousso:2009dm}. It is interesting that the model exhibits the same type of global-local duality that was found in \cite{Bousso:2009mw}.

We will conclude by mentioning a number of puzzling questions. The questions are not new: they are shared by many of the current continuum descriptions of eternal inflation.

\bi
\item The most glaring problem is that the model is not quantum mechanical. It is based on classical stochastic evolution. A quantum mechanical version of it in which decay probabilities are replaced by amplitudes, and paths are coherently summed, may be possible. This would have to be done at the level of microstates; not the macroscopic states $n,$ each of which represents $e^{S_n}$ microstates. Such a description could be made to resemble a Wheeler DeWitt formulation but it seems likely that when the $S_n$ are large, quantum interference may be quantitatively  unimportant.

\item Even if such a quantum version of the model can be formulated, it is not clear that it is really consistent with quantum mechanics. For example, the local form of the model in which we follow a single causal patch will not be information-conserving. Decoherence will happen every time a branch splits off the causal patch. That may be acceptable---perhaps even a good thing \cite{Bousso:2011up}---but it cannot be called conventional quantum mechanics.

\item  In the global picture the quantum mechanical model would also not be quantum mechanics:  the dimensionality of the Hilbert space would grow exponentially with $u$. Again, this may be acceptable but it is outside the usual rules.

\item  We have made a point of  the $p$-conformal symmetry and its similarity to the symmetry of de Sitter space. But the operational meaning of correlation functions between different causal patches is no clearer than in the continuum theory.

\item The choice of  the light-cone time variable is extremely natural in the cellular model. However, it is still possible to choose other time variables for the purpose of  cutting off the divergences and defining a measure. For example proper time can be defined along every causal patch. Equation \eqref{proper time} defines the proper time between two nodes. It also allows us to define the proper time measured from the initial condition of the cellular model. A proper time cutoff would be defined by terminating the tree along each path at the point where $\sum H_n^{-1}$ becomes equal to the cutoff value. The proper time measure would be defined by counting observations below that cutoff. When applied to the cosmological constant, the proper time cutoff leads to the well known youngness disaster. Scale-factor time \cite{DeSimone:2008bq} can also be defined here, by jumping the clock forward by $\log{H_{new}/H_{old}}$ at each node.

    As natural as the light-cone time may be, it still seems that additional principles are needed to pick it out from all the other possibilities for defining a cutoff.

\item  Finally, the role of terminals may be more complicated and interesting than we have allowed. For one thing, hats may play an important role in formulating a precise holographic quantum theory of eternal inflation \cite{Harlow:2010my}. For another, the problem of counting the number of observers in vacua with negative \cc \
     is as obscure in the cellular model as it is in  continuum theories.
\ei

In short, the cellular model is a simple and tractable model that exhibits many of the features, both good and bad, of conventional eternal inflation.

\section*{Acknowledgements}
We are grateful to a number of people for key insights. Yasuhiro Sekino showed us that the rate equations can be made symmetric if detailed balance holds. Ben Freivogel explained that the time of the cellular model is light-cone time. Raphael Bousso and Michael Salem spent many hours with us explaining light-cone time and local-global duality; Vitaly Vanchurin shared his experience with simple models of eternal inflation. We thank Brian Conrad, Edward Witten, Persi Diaconis, Susan Holmes, and Mehrdad Shahshahani for much help in understanding $p$-adic numbers and the $PGL(2, Q_p)$ symmetry of tree graphs.  We also thank Michael Dine, Renata Kallosh and Edward Witten for sharing their insights about BPS domain walls.  Finally we thank an anonymous referee for useful comments on a previous version of this work, which led to the inclusion of the discussion of the arrow of time in section 5.

Our work is supported by the Stanford Institute for Theoretical Physics and NSF Grant 0756174. DS also acknowledges the NSF under the GRF program.

\bigskip

\begin{appendix}
\section{~Light-cone Time in the Square Bubble Approximation}
This appendix reviews the key results in eternal inflation that motivate the cellular model.  The goals are to explain, in the context of Bousso's light-cone time \cite{Bousso:2009dm}, why the hyperbolic geometry of bubble nucleation can be sensibly replaced by simple square cells and to derive the rate equations governing the distribution of vacua.  These results are directly used in the main text of the paper only in the claim in section \ref{clocksect} that the natural clock of the cellular model is analogous to light-cone time.  The main points are standard results in the eternal inflation literature, but we do present a new relatively simple derivation of the relationship of light-cone time to FRW time in a thin-wall CDL bubble.\footnote{DH is especially grateful to M. Salem for a series of discussions on the issues in this appendix.  Our computation in section A.3 is closely related to his analysis with Vilenkin of the CAH+ measure in \cite{Salem:2011mj}.}
\subsection{Preliminaries}
Light-cone time is defined in the following way: begin with some initial spacelike surface $\Sigma_0$ and draw a congruence of timelike future-directed geodesics orthogonal to the surface.  Now say $P$ is an event lying in the future of $\Sigma_0$.  Consider the subset of geodesics which intersect the future lightcone of $P$.  If we call the physical volume these geodesics occupy in the initial surface $V_P$, then the ``light-cone time of $P$'' is\footnote{In this appendix we will work in 3+1 dimensions.}
\be
\label{lcdef}
u(P)=-\frac{1}{3}\log\left[(H_0)^3 V_P\right].
\ee
Here $H_0$ is some arbitrary scale, which we can for example choose to be Planck units.  We can interpret this as defining the volume at future infinity of the lightcone of $P$ as $H_0^3 V_P$ in equation \eqref{lightcone time}.  To get some intuition, consider the case of a single de Sitter space with metric
\be
ds^2=-dt^2+\frac{1}{H^2}e^{2Ht}(d\vec{x})^2.
\ee
For an initial surface choose $t=t_0$.  The congruence of geodesics are comoving.  To compute the lightcone time of a point $P$ at time $t_1$, we note that its future lightcone at time $t$ has comoving radius
\be
r=e^{-Ht_1}\left( 1-e^{-H(t-t_1)}\right),
\ee
which for $t\gg t_1$ asymptotes to $e^{-Ht_1}$.  In computing the light-cone time of $P$ we should therefore include geodesics in the comoving region $r<e^{-Ht_1}$, which has volume $\frac{4\pi}{3H^3}e^{3H(t_0-t_1)}$.  This then gives\footnote{We will not discuss scale-factor time in detail in this appendix, but we observe that this expression puts light-cone time in the same general family of slicings as scale-factor time, which would have given $u_{sf}=H(t_1-t_0)$.  The $H$-dependent shift between them leads to slightly different phenomenology when these slicings are used as cutoffs for the measure problem of eternal inflation.}
\be
u(P)=H(t_1-t_0)+\frac{1}{3}\log \frac{3H^3}{4\pi H_0^3}.
\ee

A more difficult computation is to find the light-cone time in a Coleman-De Luccia bubble.  We will work exclusively in the thin-wall approximation, for which the metric inside of the bubble is
\be
ds^2=-d\tau^2+h^{-2}\sinh^2 (h\tau) \left[d\xi^2+\sinh ^2 \xi d\Omega_2^2\right].
\ee
We tackle the geometry of the light-cone time slices inside this bubble in section \ref{lcinbubs}.  Our result, for a domain wall with zero tension, an internal Hubble $h$ which is much smaller than the ancestor Hubble $H$, and for FRW times larger than $h$, is that
\be\label{lcinbub}
u-u_{nuc}=h\tau +f(\xi),
\ee
where $f(\xi)$ is an $h$-independent function that approaches an $\mathcal{O}(1)$ constant for small $\xi$ and becomes linear at large $\xi$.  For points in the center of the bubble we also produce an exact expression that makes none of these approximations, but this result will already be sufficient for our statements in the main text of this paper.

\subsection{The Rate Equation}
\label{apprates}
In this section we derive the light-cone time rate equation \eqref{apprate} for volume measured in Hubble units.  In keeping with the literature on this subject we will make the plausible first assumption that the time scales of interest for decays are much longer than all Hubble constants, and also that the ratio of Hubble constants in any particular decay is far from unity.  Consider the change $dV_m$ in the volume of vacuum $m$ in a small step $dt$ of proper time.  The exponential expansion of dS space will produce a contribution $3H_mdt=3du$, and there will also be a positive contribution from bubbles of $m$ being nucleated in other vacua.  Similarly there will be a subtraction for bubbles of other vacua nucleating in $m$.

We first consider the negative contribution from $m$ decaying to $n$.  Say a bubble of $n$ nucleates in vacuum $m$ at time $t_{nuc}$.  From the point of view of vacuum $m$ it will rapidly expand to a comoving radius of $e^{-H_m t_{nuc}}$, after which the physical volume it occupies in $m$ will be
\be
V=\frac{4\pi}{3H_m^3}e^{3H_m(t-t_{nuc})}=\frac{4\pi}{3H_m^3}e^{3(u-u_{nuc})}.
\ee
Since we have assumed that decay rates are slow compared to $H_m$, we can approximate this process by simply removing a cube of proper volume $V_{out} \approx \frac{1}{H_m^3}$ at time $t_{nuc}$.  We drop the order one factor since we will not keep track of it in other terms.  This is called the ``square bubble approximation'' \cite{Bousso:2007nd}, and it produces a contribution to $dV_m$ of the form
\be
dV_m\supset-\sum_n \Gamma_{nm}\frac{1}{H_m^3}dt=-\sum_n \Gamma_{nm}\frac{1}{H_m^4}du.
\ee
Here $\Gamma_{nm}$ is the proper decay rate from $m\to n$.

The positive contribution to $dV_m$ coming from bubble nucleation into $m$ from other vacua is more challenging to compute.  Using equation \eqref{lcinbub} we can compute the induced metric on the cutoff surface inside the bubble
\be
ds^2\approx \frac{1}{2h^2}e^{2(u-u_{nuc}-f(\xi))}\left[(1-2 e^{-2(u-u_{nuc}-f(\xi))}f'(\xi)^2)d\xi^2+\sinh^2\xi d\Omega_2^2\right],
\ee
where we assumed that $h\tau\gg1$.  The total volume is finite, and since for $\xi>1$ we can approximate $f(\xi)\approx\xi$ it is not hard to see that almost all of the volume comes from $0<\xi<1$.  Up to an order one factor the total volume inside of the bubble of a slice of constant light-cone time $u$ is thus
\be
V=\frac{1}{H_m^3}e^{3(u-u_{nuc})}.
\ee
So in the square bubble approximation we can write
\be
dV_m\supset\sum_n\Gamma_{mn}\frac{1}{H_m^3} dt=\sum_n\Gamma_{mn}\frac{1}{H_m^3 H_n} du.
\ee
The factor of $H_n$ appears because in converting $dt$ to $du$ we need to use the ancestor Hubble since that is the appropriate proper time to multiply the decay rate by when computing differential probability to decay in time $dt$.

We can now combine these results into the light-cone rate equations for proper volume
\be
\frac{dV_m}{du}=3V_m+\sum_n\Gamma_{mn}\frac{1}{H_m^3 H_n}V_n -\sum_n \Gamma_{nm}\frac{1}{H_m^4}V_m.
\ee
We can make this look nicer by defining a dimensionless decay rate $\gamma_{mn}=\Gamma_{mn}H_n^{-4}$ and multiplying both sides by $H_m^3$ to get
\be
\label{apprate}
\frac{d(H_m^3V_m)}{du}=3(H_m^3V_m)+\sum_n\gamma_{mn}(H_n^3V_n) -\sum_n \gamma_{nm}(H_m^3V_m),
\ee
which is the light-cone time rate equation.  It is easy to see that the quantity $e^{-3u}\sum_m H_m^3 V_m$ is conserved by this equation, so the total number of horizon volumes $\sum_m H_m^3 V_m$ grows exponentially.\footnote{We have been quite cavalier in assuming that the results we derived for downward transitions are also valid for upward transitions.  This is standard practice in the literature, but that doesn't mean it is justified.  This process requires a large fluctuation in the geometry, so the definition of the time in terms of geodesics will break down and perhaps we can just take a definition of the time slicing through an upward transition that agrees with \eqref{apprate}.}

\subsection{Light-cone Time in a Bubble}
\label{lcinbubs}
In this section we derive equation \eqref{lcinbub} by computing the light-cone time for various points inside a CDL bubble.  This section is rather technical and readers who are not measure enthusiasts are welcome to skip it.  Similar computations have been done for other slicings in \cite{Vilenkin:1996ar,Bousso:2007nd,Bousso:2008hz,DeSimone:2009dq,Salem:2011mj}, but we will use geometric tricks to simplify these analyses and obtain some nice ``exact'' results.  We will assume that decay rates are small enough that the bubble nucleates well after its ancestor and thus that the light-cone time is comoving inside of the ancestor.  To compute the light-cone time of a point with FRW coordinates $(\tau,\xi)$ inside the bubble, we need to compute its future lightcone and then compare this to the fate of geodesics that begin as comoving in the ancestor.  These comoving geodesics become comoving inside the bubble at late times, and we will first compute the relationship between these two comoving distances.

Following geodesics as they pass through the domain wall is made tractable by embedding the entire geometry into a higher-dimensional Minkowski space and treating the geodesics and the domain wall as intersections of the geometry with various planes.  We will for the remainder of this section work in units where the ancestor Hubble $H$ is set equal to one.  Because of the symmetry the problem is really $1+1$ dimensional, so we will temporarily surpress two of the dimensions.  The ancestor dS space is given by the embedding
\be\label{ancemb}
-T^2+X^2+Y^2=1,
\ee
into a $2+1$-dimensional Minkowski space.  In flat slicing the ancestor has metric:\footnote{These coordinates are related to the embedding coordinates by $T=\sinh t+\frac{1}{2}e^t r^2$, $X=\cosh t-\frac{1}{2}e^t r^2$, and $Y=re^t$.}
\be
ds^2=-dt^2+e^{2t}dr^2
\ee
The bubble dS space will be a similar embedding but centered on a new origin:
\be\label{bubemb}
-T^2+(X-X_0)^2+Y^2=\frac{1}{h^2}.
\ee
For the decay to work we need $h<1$.  The domain wall is the intersection of the two embeddings with a plane
\be
X=X_w,
\ee
which allows us to determine
\be
X_0=X_w-\sqrt{X_w^2+\frac{1}{h^2}-1}.
\ee

The domain wall is of course not a geodesic, but it is a convenient fact that the intersection of a plane that passes through the origin of the embedding space with the surface \eqref{ancemb} \textit{is} a geodesic in the induced metric on the ancestor dS space.  Similarly the intersection of a plane that passes through the point $T=Y=0,X=X_0$ with the surface \eqref{bubemb} is also a geodesic inside the bubble.  We can construct a geodesic that passes from the ancestor into the bubble by finding two planes, one passing through the origin and one passing through $T=Y=0,X=X_0$, which both intersect the domain wall at the same point and whose induced geodesics have tangent vectors that have the same angle with the domain wall at that point.\footnote{We thank M. Salem and I. Yang for pointing out a problem with an earlier version of this argument.}  These two planes can be parameterized as
\begin{align}\nonumber
-\hat{A}T+\hat{B}X+Y=&0\\
-AT+B(X-X_0)+Y=&0.
\end{align}
$\hat{A}, \hat{B}$ are determined in terms of the comoving radius $r_0$ of the geodesic in the ancestor by
\be\hat{A}=-\hat{B}=r_0.
\ee
$A$ and $B$ determine the trajectory of the geodesic inside of the bubble in FRW coordinates, via\footnote{Here the relationship of the embedding coordinates to the FRW coordinates is $T=\sinh \tau \cosh \xi$, $Y=\sinh \tau \sinh \xi$, and  $X-X_0=\cosh \tau$.}
\be
-A\sinh(h\tau)\cosh \xi+B\cosh (h\tau)+\sinh(h\tau) \sinh \xi=0.
\ee
At late times in the bubble this becomes
\be
\label{frwgeod}
A\cosh \xi=B+\sinh \xi,
\ee
which is manifestly comoving at location $\xi$.  $A$ and $B$ are determined in terms of $r_0$ by the following steps: solve
\begin{align}\nonumber
-T_w^2+X_w^2+Y_w^2=&1\\
r_0(T_w+X_w)-Y_w=&0\label{wallplace}
\end{align}
to determine the location of the intersection of the geodesic with with the domain wall, solve
\begin{align}\nonumber
-r_0V_T-r_0V_X+V_Y=&0\\
-T_wV_T+X_w V_X+Y_wV_Y=&0 \label{tangentvec}
\end{align}
to determine the tangent vector $(V_T,V_X,V_Y)$ to the geodesic at the intersection outside of the bubble, solve\footnote{The second of these two equations imposes that the tangent vector has the same angle with the geodesic on both sides; we have chosen the norms of $V$ and $V'$ to be equal.}
\begin{align}\nonumber
-T_w V_T'+X_wV_X'+Y_wV_Y'=&0\\
-V_T+\frac{T_w}{Y_w}V_X=&-V_T'+\frac{T_w}{Y_w}V_X'\label{intangentvec}
\end{align}
to determine the tangent vector $(V_T',V_X',V_Y')$ to the geodesic at the intersection inside of the bubble, and finally solve
\begin{align}\nonumber
-AT_w+B(X_w-X_0)+Y_w=&0\\
-AV_T'+B V_X'+V_Y'=&0 \label{ABeq}
\end{align}
to find $A$ and $B$.  These equations are at most quadratic, so it is straightforward to write down the exact solution.  The result however is somewhat unwieldy, simpler expressions are available if we work in the limit of zero tension in the domain wall, for which $X_w=1$:
\begin{align}\nonumber
A&=\frac{r_0(1-r_0(1+h^2)/2)}{1-r_0(1-r_0(1-h^2)/2)}\\
B&=-\frac{hr_0(1-r_0)}{1-r_0(1-r_0(1-h^2)/2)}.
\end{align}
Combining these with \eqref{frwgeod} gives the desired relationship between the comoving locations of the geodesic inside and outside of the bubble.  Athough this result was computed in 1+1 dimensions it is valid in any dimension by symmetry.

We now compute the lightcone time for a point $P$ in the center of the bubble with FRW time $\tau$.  The future lightcone of $P$ is easy to compute, it has asymptotic comoving radius
\be
\xi=-\log \left(\tanh(h\tau/2)\right).
\ee
Using this in \eqref{frwgeod} we find
\be
\label{taur}
A\cosh(h\tau)=B\sinh (h\tau)+1.
\ee
This expression gives the exact relationship between the FRW time $\tau$ and the ancestor comoving radius $r_0$ inside of which geodesics enter the future lightcone of $P$.  Finally the light-cone time \eqref{lcdef} of $P$ is given by
\be
\label{lctimer}
u(P)=u_{nuc}-\log r_0.
\ee
We have given exact prescriptions for computing $A$ and $B$ in terms of $r_0$, but it is convenient to give the result in the special case where $h\tau \gg 1$, $h\ll 1$ (remember $H=1$), and the domain wall tension is zero.  These approximations are justified in the beginning of section \ref{apprates}, and they lead to $r_0\ll1$.  We then simply have $A\approx r_0$ and $B\approx -h r_0$, so from \eqref{taur} we find that the lightcone time of $P$ in this approximation is
\be
\label{lccenter}
u(P)\approx u_{nuc}+ h\tau-\log 2.
\ee
The corrections to this formula are parametrically small in $h/H$ and $h\tau$.

Computing the light-cone time of points with $\xi\neq0$ is more challenging, we will only sketch a computation of the asymptotic behavior for large $\xi$ with the same approximations.  Say that point the $P$ we are interested is at $(\tau_0,\xi_0)$.  Its future light-cone is no longer centered at $\xi=0$, but we note that the null geodesics starting at $P$ and moving inwards and outwards radially asymptote to $\xi=\xi_0\mp \log\left[\tanh(h\tau_0/2)\right]$.  For large $h\tau_0$ we can write this as:
\be
\xi=\xi_0\pm2e^{-h\tau_0}+\mathcal{O}(e^{-2h\tau_0}).
\ee
Feeding this into \eqref{frwgeod} we find that these two geodesics have comoving radii
\be
r_{\pm}=1-\sqrt{2}e^{-\xi_0}(1\mp2e^{-h\tau_0}).
\ee
We can then approximate the volume in the initial surface as a sphere whose radius is given by
\be
r_0\approx \frac{r_+-r_-}{2}=2\sqrt{2}e^{-\xi_0-h\tau_0}.
\ee
This result is accurate up to a multiplicative order one factor.  Finally using this in \eqref{lctimer} we find
\be
u(P)=u_{nuc}+h\tau_0+\xi_0+\mathcal{O}(1).
\ee
This expression, along with \eqref{lccenter}, completes our derivation of \eqref{lcinbub}.

\section{ dS/CFT and FRW/CFT}
dS/CFT and FRW/CFT are two holographic proposals for dual theories of eternal inflation.  In this appendix we make some preliminary remarks on the implications for these theories of our results in the cellular model.
\subsection{dS/CFT}
The term ``dS/CFT'' is applied to several different closely related ideas \cite{Witten:2001kn, Strominger:2001pn, Maldacena:2002vr}: the most precise is the proposal that the Wheeler-deWitt wave function of de Sitter space is computed by the partition function of a nonunitary Euclidean conformal field theory deformed by various sources which are simply related to the arguments of the wave function.\footnote{An explicit example of this was recently proposed in Vasiliev gravity in \cite{Anninos:2011ui}.} This wave function does not have a simple analogue in the stochastic model we have described in this paper.  Our model is not quantum mechanical, and the probability distribution for colors at future infinity we study is more analogous to the semiclassical limit of the square of the wave function.  The $PGL(2,Q_p)$ symmetry we found in the model thus does not directly parallel the conformal invariance of the CFT that computes the wave function.  In  \cite{Harlow:2010my, Harlow:2011ke} the idea was explored that the square of the wave function can be interpreted as a partial integration of a new ``doubled'' CFT living at future infinity, which includes as dynamical fields two copies of the fields from the wave function CFT and also the boundary values of the bulk fields.  In \cite{Harlow:2011ke}  it was shown that it is this doubled CFT whose correlators compute the extrapolation of bulk correlators to the future boundary.  Since the correlators we studied in this paper were first computed on the tree and then extrapolated to the boundary, they are analogous to correlators in the doubled CFT.

Bubble nucleations are not well-understood in dS/CFT.  In \cite{Freivogel:2009rf} evidence was provided that in a landscape without terminals the correlation functions  of bubbles at future infinity should be conformally invariant.  Our work confirms that this is so in the cellular model, which as we just discussed has a boundary theory analogous to the doubled CFT in dS/CFT.  The multiverse fields of the cellular model suggest that there is a large sector of highly relevant operators in the doubled CFT whose dimensions are very close to zero and whose correlators describe the bubble distribution at future infinity.  It would be good to understand how to express these operators in terms of the doubled CFT fields.

The situation is more interesting when terminals are included.  In the cellular model we found that the correlators of operators at points that were conditioned to not be terminal had an interesting nonconformal structure we referred to as a fractal flow.  An interesting conjecture is that this is also what happens in the doubled CFT in the continuum theory.\footnote{Another interesting conjecture for including terminals is discussed in  \cite{Garriga:2008ks, Garriga:2009hy, Garriga:2010fu} .}  This is a surprisingly simple proposal that we hope to return to in future work.

\subsection{FRW/CFT}
\begin{figure}[h]
\begin{center}
\includegraphics[scale=.4]{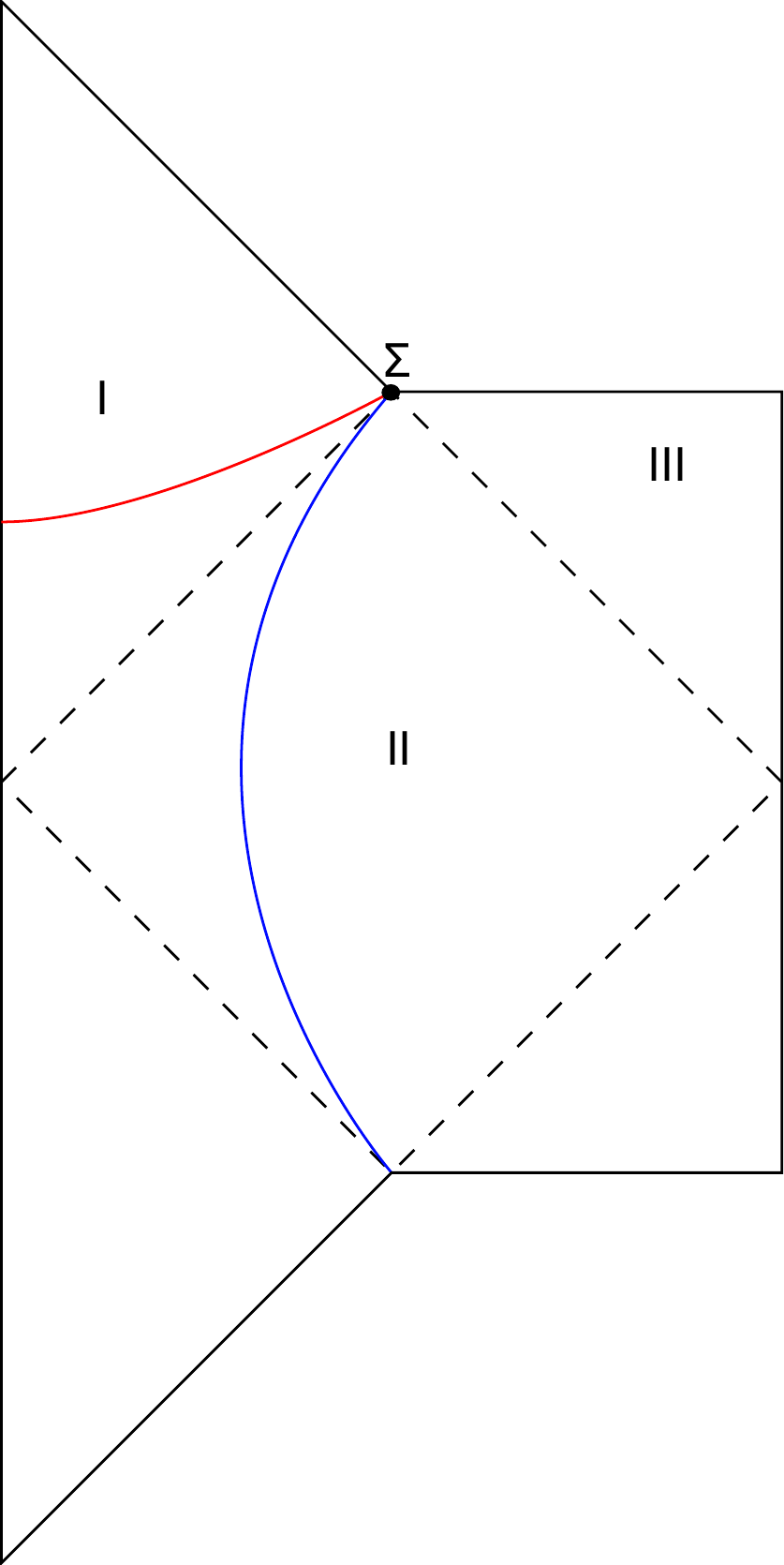}
\caption{The Penrose diagram of the Lorentzian Coleman De-Luccia geometry.  The dashed lines separate it into different regions which are preserved by symmetry.  The lower half is the time-reversal of the upper half and is unphysical.  Region I is inside of the bubble and the red line is a slice of constant FRW time.  Region II is the domain wall, which is foliated by lower-dimensional dS slices.  A representative dS slice is shown in blue.  Region III is the future dS boundary.  The point $\Sigma$ is the two-sphere where the FRW/CFT theory lives.  It can be thought of either as a spatial boundary of region I or a future boundary of region II.  Since the natural spatial slices of region II have finite volume we can dimensionally reduce to a single lower dimensional dS theory, whose future boundary is $\Sigma$.  It is clear that a timelike observer in region I, a census-taker, will see $\Sigma$ at late times.}
\label{hat1}
\end{center}
\end{figure}

 FRW/CFT proposes that correlation functions seen on the sky of an observer living in a stable $\Lambda=0$ bubble, usually called the Census Taker, are computed by a Euclidean CFT living on a sphere.  It was argued in \cite{Harlow:2010my} that this dual theory is closely related to a dimensionally reduced dS/CFT theory whose correlators are computed by extrapolating bulk correlators in the domain wall region of the CdL geometry to the future boundary.  This is illustrated in figure \eqref{hat1}.

Bubbles that are nucleated within the domain wall region, denoted II in figure \eqref{hat1}, are inside the horizon of the Census Taker, and they produce a pattern of circles on his/her sky which is simply related to the distribution at future infinity of the dimensionally reduced dS theory\footnote{Early ideas about describing bubble collisions in FRW/CFT by a lower dimensional effective theory  were developed as one aspect of unpublished work on the ``Census Taker Measure" by Raphael Bousso Ben Freivogel, Alex Maloney, Stephen Shenker, Leonard Susskind and I-Sheng Yang.}.  This lower dimensional theory is gravitational, that is  it has a normalizable Randall Sundrum type graviton mode as in the dS/dS correspondence \cite{Alishahiha:2004md}.   So we expect it to be able to  make both up and down transitions,   and that unless there are lower-dimensional terminals it will be conformally invariant.   The only candidates for terminals are nucleations of $\Lambda < 0$ bubbles in region II.  This situation has been analyzed in \cite{Freivogel:2007fx}.   If the domain wall has high tension it accelerates out of the Census Taker bubble.  Here the domain wall geometry is that of a lower dimensional dS space  that is essentially region II including its gravitational character.   Its ability to nucleate up and down transitions means that it is not a terminal.  The only exceptions are   $\Lambda <0$  bubbles whose domain walls with the $\Lambda=0$ region have exactly BPS tension.    Here the domain wall geometry is flat and no up or down transitions are possible.   It serves as a ``hat" terminal in the lower dimensional dS space of region II.   These BPS domain walls are terminals from the point of view of the bubble distribution seen by the Census Taker, since the circles they produce do not have any smaller structure inside of them.

The existence of BPS domain walls in the string landscape is an open problem, but  the fragmentary evidence available seems to suggest that generically they do exist.\footnote{We thank Michael Dine, Renata Kallosh and Edward Witten for discussions of this point.}
 Our analysis of the cellular model with terminals clearly has implications for how BPS domain walls might be incorporated into FRW/CFT, and we hope to return to this idea in more detail in the future.

\end{appendix}

\end{document}